\documentclass[bibyear]{aa}  
\usepackage[T1]{fontenc}

\makeatletter
\renewcommand*\aa@pageof{, page \thepage{} of \pageref*{LastPage}}
\makeatother

\usepackage{graphicx}
\usepackage{txfonts}
\usepackage{natbib}
\bibpunct{(}{)}{;}{a}{}{,} 
\usepackage{booktabs}
\usepackage{multicol}
\usepackage{graphicx}
\usepackage{fancyhdr}
\usepackage[hidelinks]{hyperref}
\usepackage{amsmath}	
\usepackage{amssymb}	
\usepackage[inter-unit-product=\cdot]{siunitx}
\usepackage{enumerate}
\usepackage{multirow}
\usepackage{mathtools}
\usepackage{comment}
\usepackage{longtable}
\usepackage{tabularx}
\usepackage{xcolor} 
\usepackage{ulem} 
\usepackage{comment}
\usepackage{cuted}
\usepackage{color, soul}
\usepackage{array}
\usepackage{graphicx}
\usepackage{txfonts}
\usepackage{subcaption} 

\newcommand{\orcidicon}[1]{\href{https://orcid.org/#1}{\includegraphics[width=11pt]{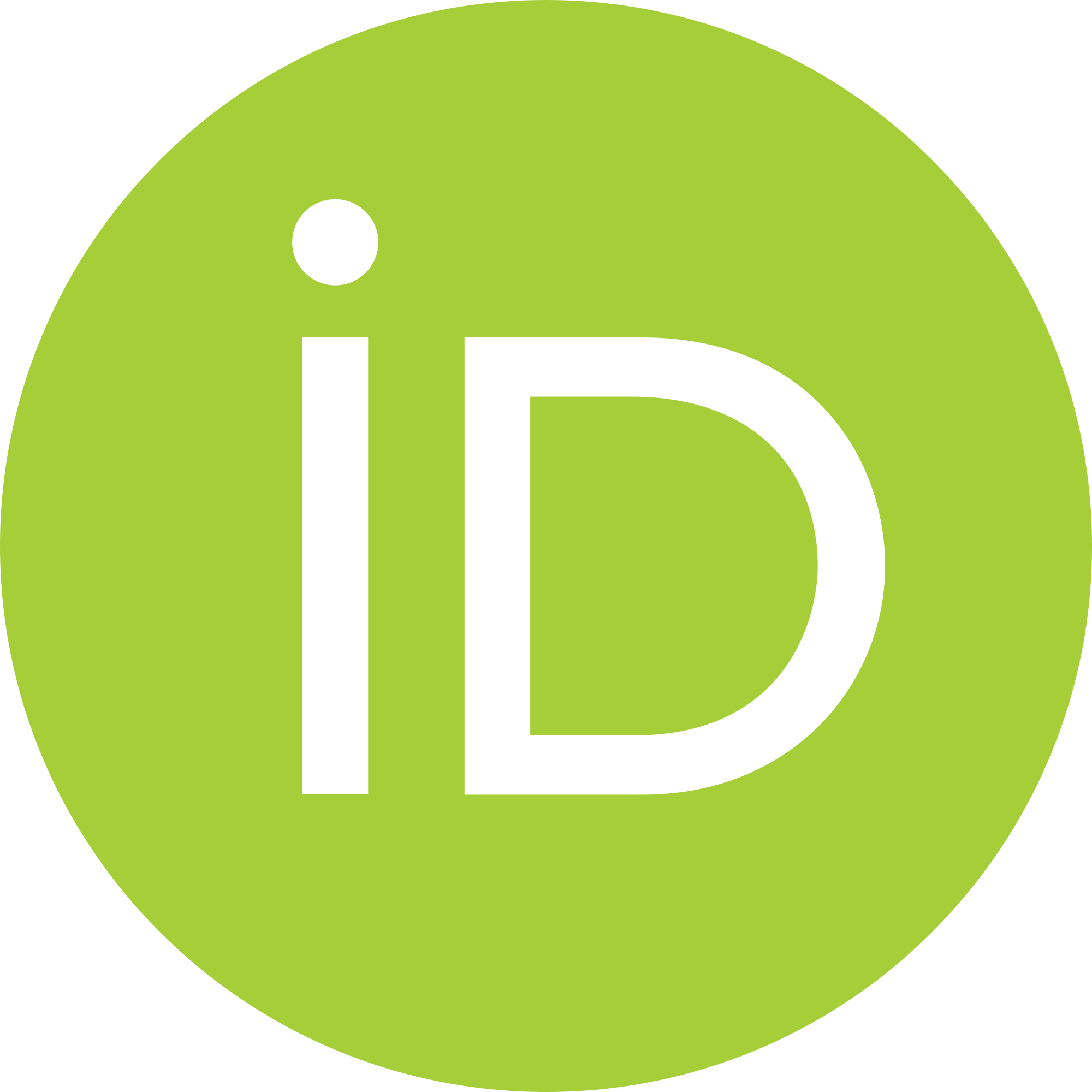}}}
\newcommand{\orcid}[1]{\href{https://orcid.org/#1}{\protect\orcidicon{#1}}}

\begin{document}

   \title{Eccentric black hole mergers via three-body interactions in young, globular, and nuclear star clusters}
   \titlerunning{Eccentric BH mergers via three-body interactions}


   \author{
    Marco Dall'Amico,\inst{1,2}\orcid{0000-0003-0757-8334}\thanks{\href{mailto:marco.fromfriend@gmail.com}{marco.dallamico@phd.unipd.it}}
    Michela Mapelli,\inst{3,1,2,4}\orcid{0000-0001-8799-2548}\thanks{\href{mailto:mapelli@uni-heidelberg.de}{mapelli@uni-heidelberg.de}}
    Stefano Torniamenti\inst{3,1,2,4}\orcid{0000-0002-9499-1022}
    \and Manuel Arca Sedda\inst{5,1,2}\orcid{0000-0002-3987-0519}}
    \authorrunning{Marco Dall'Amico et al.}
    \institute{
    $^{1}$Physics and Astronomy Department Galileo Galilei, University of Padova, Vicolo dell’Osservatorio 3, I–35122, Padova, Italy\\
    $^{2}$INFN-Padova, Via Marzolo 8, I–35131 Padova, Italy\\
    $^{3}$Institut f{\"u}r Theoretische Astrophysik, ZAH, Universit{\"a}t Heidelberg, Albert-Ueberle-Str.~2, D-69120, Heidelberg, Germany\\
    $^{4}$INAF–Osservatorio Astronomico di Padova, Vicolo dell’Osservatorio 5, I–35122, Padova, Italy\\
    $^{5}$ Gran Sasso Science Institute (GSSI), 67100 L’Aquila, Italy}

   \date{Received November 21, 2023}

 
  \abstract{
    Eccentric mergers are a signature of the dynamical formation channel of binary black holes (BBHs) in dense stellar environments and hierarchical triple systems. Here, we investigate the formation of eccentric mergers via binary-single interactions by means of $2.5\times10^{5}$ direct \textit{N}-body simulations. Our simulations include post-Newtonian terms up to the 2.5th order and model the typical environment of young (YSCs), globular (GCs), and nuclear star clusters (NSCs). Around $0.6\%$ ($1\%$) of our mergers in NSCs (GCs) have an eccentricity ${>0.1}$ when the emitted gravitational wave frequency is 10 Hz in the source frame, while in YSCs this fraction rises to $1.6\%$.  Approximately $\sim63\%$ of these mergers are produced by chaotic, resonant interactions where temporary binaries are continuously formed and destroyed, while $\sim31\%$ arise from an almost direct collision of two black holes (BHs).
    Lastly, $\sim 6\%$ of these eccentric mergers occur in temporary hierarchical triples. We find that binaries undergoing a flyby generally develop smaller tilt angles with respect to exchanges. This result challenges the idea that perfectly isotropic spin orientations are produced by dynamics. The environment dramatically affects BH retention: $0\%$, $3.1\%$, and $19.9\%$ of all the remnant BHs remain in YSCs, GCs, and NSCs, respectively. The fraction of massive BHs also depends on the host cluster properties, with pair-instability ($60\leq\,$M$_{\rm BH}$/M$_{\odot}\leq$100) and intermediate-mass (M$_{\rm BH}\geq$100$\,$M$_{\odot}$) BHs accounting for approximately $\sim44\%$ and $1.6\%$ of the mergers in YSCs, $\sim33\%$ and $0.7\%$ in GCs, and $\sim28\%$ and $0.4\%$ in NSCs, respectively.}


   \maketitle
%

\section{Introduction}

Binary-single encounters dominate the interactions between black holes (BHs) in the core of star clusters \citep{heggie1975,hut1983,Hut1983b,Hut1993,Banerjee2010}. In this region, BHs form a dynamically decoupled sub-core where they can mostly interact via binary-single scatter due to the large cross-section of this process \citep[e.g.][]{breen2013a,breen2013b,samsing2014}. This 
springs from star clusters' tendency to evolve towards energy equipartition \citep[][]{spitzer1969,trenti2014,spera2016,bianchini2016}, combined with dynamical friction acting on the most massive bodies of the cluster \citep[]{Meylan1997,Fregau2002,Gurkan2004}.

Most binary black holes (BBHs) in star clusters belong to the family of hard binaries; that is, binary systems with a binding energy larger than the average star kinetic energy of the cluster. Since hard binaries statistically tend to get harder during binary-single encounters \citep{heggie1975}, BBHs tend to progressively decrease their semi-major axis, or even increase their total mass if a dynamical exchange with a single BH takes place \citep{hillsfullerton1980}. 
Three-body interactions\footnote{Hereafter, we will use the terms three-body interactions and binary-single encounters as synonyms.} between BHs are therefore a key mechanism to speed up the merger before gravitational wave (GW) emission becomes efficient. These dynamical encounters not only efficiently produce BBH mergers \citep[e.g.][]{Sigurdsson1995,portegieszwart2000,Banerjee2010,tanikawa2013,ziosi2014,Morscher2015,Rodriguez2015,Rodriguez2016a,Rodriguez2016b,mapelli2016,samsing2017,samsing2018b,samsingilian2018,Trani2019,Trani2021,dallamico2021} and BH-neutron star mergers \citep[e.g.][]{2013MNRAS.428.3618C, 2020CmPhy...3...43A, 2021ApJ...908L..38A}, but they may also cause eccentric mergers \citep{Gultekin2006,samsing2014,Samsing2017b,samsing2018a,Samsing2018c,Rodriguez2018,Zevin2019,2021A&A...650A.189A,trani2022,codazzo2023}. These are mergers in which the coalescence time of the binary is shorter than the timescale it takes for GW emission to circularise the orbit, such that the binary can merge with a non-zero eccentricity in the LIGO--Virgo sensitivity band \citep{abbotteccentric}. In dynamical interactions, the energy exchange between the bodies can excite the eccentricity of a BBH and even induce it to merge. This is particularly true for three-body interactions, in which the system can evolve into a chaotic regime with temporary binaries that are continuously created and destroyed. In this unstable triple configuration, the single BH can perturb the temporary binary and induce it to merge rapidly enough that GWs are not sufficient to completely circularise the orbit \citep{samsing2014}.

Isolated binaries, on the other hand, struggle to produce BBHs with a non-negligible eccentricity at merger. Tidal effects, mass transfer episodes, and common envelope events usually circularise the orbit of a binary star even before it evolves into a BBH \citep{hurley2002}. Even if supernova kicks can increase the eccentricity of the system, GWs efficiently circularise the orbits by the merger time \citep{peters1964}. Eccentric mergers are therefore commonly associated with BBHs formed in a dynamically active environment. Eccentricity, if detected in the waveform of a merger, might be used as a tool to infer the dynamical origin of a BBH \citep[][]{Amaro2016,Chen2017,Gayathri2020,Romero2020,Romero2021,Zevin2021}.

How often are these eccentric mergers produced by dynamical interactions? And in which environment should we expect them to be more frequent? Here, we aim to address these questions via direct \textit{N}-body simulations of three-body encounters between BBHs and BHs. We performed three different sets of 
binary-single scattering experiments, each with different initial conditions appositely set to reproduce the properties of a class of star clusters: young star clusters (YSCs), globular clusters (GCs), and nuclear star clusters (NSCs). Our goal is to investigate the effect of the cluster properties on the interactions and to derive the influence that the hosting environment has on the outcomes and production of eccentric BBH mergers\footnote{The data underlying this article are available at the following Zenodo link: \url{https://doi.org/10.5281/zenodo.7684085}}.

Three-body interactions are the fundamental mechanism at the base of hierarchical mergers; in other words, the process in which two BHs merge and their merger remnant 
collides with other BHs of the cluster, giving rise to multiple generations of BBHs \citep[][]{miller2002,Gerosa2017,Fishbach2017,Rodriguez2019,Antonini2019,Doctor2020,Arcasedda2021a,mapelli2021a,Gerosa2021,Atallah2022}. 
Here, we discuss the impact of three-body recoil velocities on hierarchical mergers, and how this effect, combined with relativistic kicks and star cluster evaporation, could dynamically eject the BHs from the cluster.


\section{Methods}


\begin{figure*}
	\includegraphics[width=2.0\columnwidth]{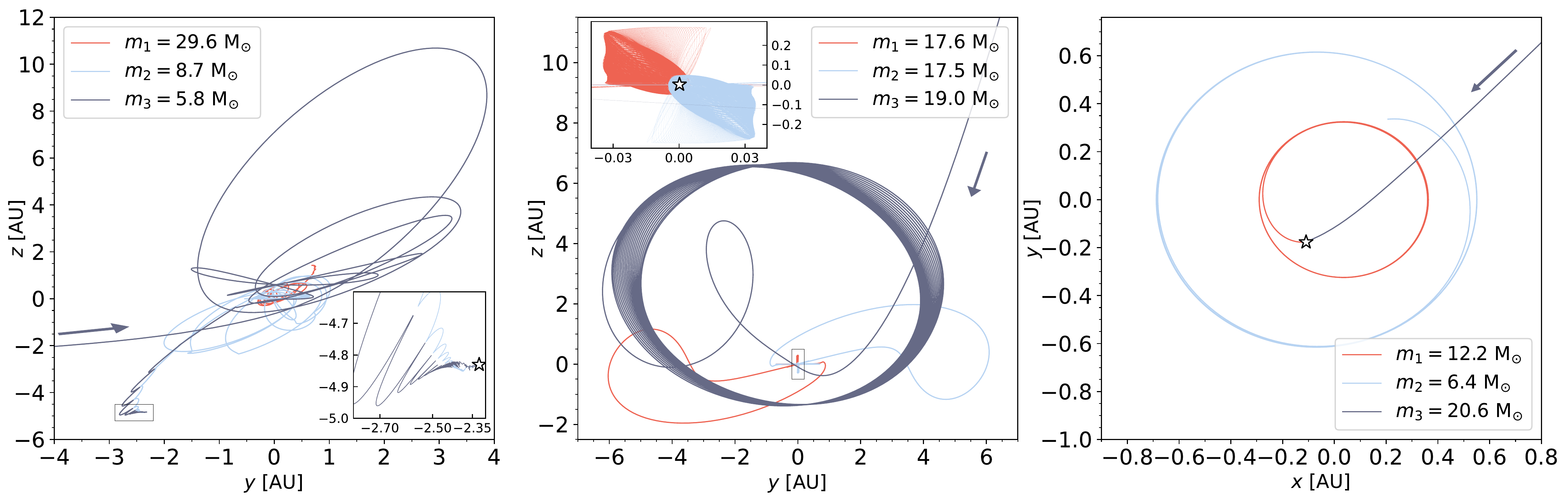}
    \caption{Trajectories of three 
    eccentric mergers triggered by a three-body interaction in NSCs (left- and right-hand panels) and GCs (central panel). In the three panels, the systems are centred in the centre of mass of the initial BBH, while the position at which the merger takes place is marked with a white star. The trajectories of the primary and secondary BH of the initial BBH (with mass $m_{1}$ and $m_2$), and the intruder ($m_{3}$)  are reported 
    in red, blue, and grey, respectively. The incoming direction of the intruder is shown with an arrow. The insets in the left-hand and central panels 
    display a close-up view of the merger region. 
    In the left-hand and central panels, the initial binary is shown edge-on, while in the right-hand panel the view is face-on. The initial coalescence time of these BBHs, i.e. the coalescence time of the initial binary at the beginning of the interaction, is longer than the Hubble time.}
    \label{fig:traj}
\end{figure*}

\subsection{Direct \textit{N}-body simulations}

Three-body encounters are chaotic processes \citep{Poincare,Valtonen}. Due to this nature, the orbits of the interacting bodies are highly unpredictable, and no general analytical solutions are known to exist. Potentially, even a perturbation of the Planck-scale order applied to the initial conditions can exponentially grow and lead to different final configurations of the system \citep{samsingilian2018,Manwadkar2020,boekholt2020,boekholt2021,pzwart2021,Hugo2021}. Therefore, the most convenient approach to studying the three-body problem from the perspective of BBH mergers is to use a numerical integrator over a large set of interactions and derive the statistical properties of the encounters.

Here, we used the direct \textit{N}-body code {\sc arwv} to simulate three different sets of three-body interactions, for a total of $2.5\times10^{5}$ simulations between a BBH and a single BH. Each of our three sets of simulations was initialised with different initial conditions, designed to reproduce the properties of three-body encounters that take place inside YSCs, GCs, and NSCs. {\sc arwv} is an algorithmic regularisation direct \textit{N}-body code \citep{mikkola1989,mikkola1993,arcasedda2019,chassonnery2019,chassonnery2021} that solves the equations of motion of the interacting bodies with post-Newtonian corrections up to the 2.5 order \citep{mikkola2008,memmesheimer2004}. 

We integrated each system with {\sc arwv} for a minimum time of $10^{5}$ yr. We stopped the integration at this time only if at least one merger took place, or if the 
three-body encounter was over. If none of these two conditions was satisfied (e.g. the three bodies were still interacting),
we carried on the integration with {\sc arwv} for a longer time. We stopped the simulation if at least one of the two aforementioned conditions was fulfilled, or if the time reached a maximum of 1~Myr. 
Figure~\ref{fig:traj} shows three examples of our three-body simulations computed with {\sc arwv}.

Most BBH mergers did not take place during the simulation with {\sc arwv}. Therefore, we evolved the remaining binary population 
according to \cite{peters1964}:
\begin{eqnarray}\label{eq:peters}
 \frac{{\rm d}a}{{\rm d}t}=-\frac{64}{5}\,{} \frac{G^3 \,{} m_{i} \,{} m_{j} \,{} (m_{i}+m_{j})}{c^5 \,{} a^3\,{} (1-e^2)^{7/2}}\,{}f_1(e), \nonumber\\
  \frac{{\rm d}e}{{\rm d}t}=-\frac{304}{15}\,{} e \frac{ G^3 \,{} m_{i} \,{} m_{j} \,{} (m_{i}+m_{j})}{c^5 \,{}a^4 \,{}  (1-e^2)^{5/2}}\,{}f_2(e),
\end{eqnarray}
where
\begin{eqnarray}
f_1(e)=\left(1+\frac{73}{24}\,{}e^2+\frac{37}{96}\,{} e^4\right) \nonumber\\
f_2(e)=\left(1+\frac{121}{304} \,{} e^2\right).
\end{eqnarray}
Here, $G$ is the gravity constant, $c$ the speed of light, $m_{i}$ the primary mass, $m_{j}$ the secondary mass, $a$ the semi-major axis, and $e$ the orbital eccentricity.
We assumed that two BHs merge when their mutual distance is less than the sum of their innermost stable circular orbits for non-spinning BHs; that is, when $r\leq{}6\,{}G(m_{i}+m_{j})/c^{2}$. All the binaries that survive after the end of the three-body integration with {\sc arw} are  long-lived and circularise their orbit through GW emission, unaffected by external perturbations. Hence, we can treat them with the simplified formalism described in Eqs.~\ref{eq:peters}, without the need for a more computationally expensive post-Newtonian formalism.

When two BHs merged, we implemented the relativistic fitting equations reported in \cite{Lousto2012} to compute the relativistic kick exerted on the BH remnant by the anisotropic GW emission at merger. These are:
\begin{equation}
    v_{\rm kick} = (v^{2}_{\rm m}+v^{2}_{\perp}+2v_{\rm m}v_{\perp}+\cos\phi+v^{2}_{\parallel})^{1/2},
\end{equation}
where
\begin{eqnarray}
    v_{\rm m} = A\,\eta^{2}\,\frac{1-q}{1+q}\,(1+B\,\eta)\\
    v_{\perp} = H\,\frac{\eta^{2}}{(1+q)}\, \left\lvert \chi_{\rm 1 \parallel}-q\,\chi_{\rm 2 \parallel}\right\lvert\\
    v_{\parallel} = \frac{16\eta^{2}}{(1+q)}\,\bigg[V_{\rm 1,1}+V_{\rm A}\,S_{\parallel} + V_{\rm B}\,S^{2}_{\parallel} + V_{\rm C}\,S^{3}_{\parallel} \bigg]\\ \left\lvert\chi_{\rm 1 \perp}-q\,\chi_{\rm 2 \perp}\right\lvert\,\cos(\phi_{\Delta}-\phi).
\end{eqnarray}
Here, $q=m_{\rm2}/m_{\rm1}$ with $m_{\rm2}<m_{\rm1}$, $\eta=q\,(1+q)^{-2}$, $A=1.2\times10^{4}$ km s$^{-1}$, $B=-0.93$, $H=6.9\times10^{3}\,$ km s$^{-1}$, $V_{\rm 1,1}=3678$ km s$^{-1}$, $V_{\rm A}=2481$ km s$^{-1}$, $V_{\rm B}=1792$ km s$^{-1}$, and $V_{\rm C}=1506$ km s$^{-1}$. The equations incorporate the components of the dimensionless spin parameter vectors, $\vec{\chi}_{\rm 1}$ and $\vec{\chi}_{\rm 2}$, relative to the primary and secondary BHs. Specifically, $\chi_{\rm 1\parallel}$ and $\chi_{\rm 2\parallel}$ are the primary and secondary dimensionless spin components parallel to the orbital angular momentum of the binary, while $\chi_{\rm 1\perp}$ and $\chi_{\rm 2\perp}$ are the perpendicular spin components lying in the orbital plane. Furthermore, $S_{\parallel}$ is the component parallel to the orbital angular momentum of the vector $\vec{S}=2\,(\vec{\chi}_{\rm 1}+q^{2}\,\vec{\chi}_{\rm 2})/(1+q)^{2}$. Moreover, $\phi_{\Delta}$ is the angle between the direction of the infall at merger and the in-plane component of the vector $\vec{\Delta}=(m_{\rm1}+m_{\rm2})^{2}\,(\vec{\chi}_{\rm 1}+q\,\vec{\chi}_{\rm 2})/(1+q)$, while $\phi$ is the phase of the binary. For further details, we refer to \cite{Lousto2012}. Finally, we computed the mass of the BH remnant and the magnitude of its spin with the fitting equations presented by  \cite{jimenez2017}.


\subsection{Initial conditions}\label{sec:IC}

\begin{table}
	\begin{center}
	\caption{Initial conditions.}
	\label{tab:IC}
	\begin{tabular}{lcc} 
		\hline
		     Property    & Initial distribution  & Interval \\
        \hline
        
        $m_1\,$, $m_2\,$, $m_3$ & Population~synthesis & [$5$, $60$]$\,$M$_{\odot}$  \\
        $a$ & Log-normal Distribution & $[{\rm max}(a_{\rm ej},a_{\rm gw}),a_{\rm hard}]$  \\
        $e$ & Thermal distribution & $[0,\,{}1]$  \\
        $f$ & \cite{hut1983} & $[-\pi{},\,{}\pi]$\\
        $v_{\infty}$ & Maxwellian  distribution & --\\
        $b$ & Uniform in $b^{2}$ & $[0,\,{}b_{\rm max}]$\\
        $D$ & $10^{3}\,{} a$ & --\\
        $\psi$ & Uniform & $[0,\,{}2\pi]$\\
        $\theta$ & Uniform in $\cos{\theta}$ & $[-1,\,{}1]$\\
        $\phi$ & Uniform & $[0,\,{}2\pi]$\\
        $\chi_{1}\,$, $\chi_{2}\,$, $\chi_{3}$ & Maxwellian distribution & $[0,\,{}1]$\\
        $\vec{\chi}_{1}\,$, $\vec{\chi}_{2}\,$, $\vec{\chi}_{3}$ & Isotropic spin orientation & --\\
		\hline
	\end{tabular}
	\end{center}
		\flushleft
\footnotesize{Column 1, initial conditions: Mass of the primary and secondary BH in the initial binary ($m_1$ and $m_{2}$), mass of the single BH ($m_{3}$), semi-major axis ($a$), orbital eccentricity ($e$), orbital phase of the binary ($f$),  initial relative velocity between the single BH and the BBH ($v_{\infty}$), impact parameter ($b$), initial distance ($D$) of the intruder from the binary centre of mass, three directional angles of the interaction ($\psi,\,{}\theta$, and $\phi$),  spin magnitude ($\chi_{1}$, $\chi_{2}$, and $\chi_{3}$), and direction of the three BHs ($\vec{\chi}_{1}$, $\vec{\chi}_{2}$, and $\vec{\chi}_{3}$). Column 2, the distribution that we used to sample the initial conditions. For the masses, we used the output of the population synthesis code {\sc mobse} \protect\citep{mapelli2017,giacobbo2018a}.  Column 3, the interval that we considered for each distribution. }
\end{table}

Simulating a three-body interaction with spinning BHs requires 21 initial parameters for each simulation. Since covering a $21-$dimension parameter space with direct \textit{N}-body simulations is computationally prohibitive, we initialised three different sets of $5\times{}10^4$, $10^5$, and $10^5$ simulations for 
YSCs, GCs, and NSCs, respectively. 
Table~\ref{tab:IC} reports all the initial conditions, the distribution used to generate each of them, and the interval in which the parameters are sampled. We initialised the properties of each encounter with the same procedure as in \cite{dallamico2021}, except for the masses, semi-major axis, and initial intruder velocity. 
Here, we summarise the main features of our initial conditions.

We assumed that the initial BH mass distribution 
does not change if the cluster is young, globular, or nuclear \citep{mapelli2021a}, and sampled the mass of all our single and binary BHs from a catalogue of synthetic BHs generated with 
the population synthesis code {\sc mobse} \citep{mapelli2017,giacobbo2018a,giacobbo2018b}. With this method, our population is composed of first-generation BHs produced by the evolution of binary stars at metallicity $Z=0.1\,{}Z_{\odot}$, with $Z_{\odot}=0.02$. {\sc mobse} implements up-to-date wind models for massive stars \citep{vink2001,chen2015}, the core-collapse supernova models by \cite{fryer2012} and the (pulsational) pair-instability supernova treatment presented in \cite{mapelli2020}. We  adopted 
the rapid core-collapse supernova model by \cite{fryer2012}. With this choice, our BH mass spectrum ranges between $5$ and  $60\,$M$_{\odot}$ BHs. 

The semi-major axis, $a$, of the initial binaries was sampled from a log-normal distribution as
\begin{eqnarray}
    p(a)=\frac{1}{\sigma_{\rm log}
    \sqrt{2\pi}}\,{}\exp{\left[{-\frac{(\log{a}-\mu_{\rm log})^{2}}{2\sigma_{\rm log}^{2}}}\right]}
\end{eqnarray}

with limits $[{\rm max}(a_{\rm ej},a_{\rm gw}),a_{\rm hard}]$, where
\begin{eqnarray}\label{eq:smalimits}
 a_{\rm hard}=\frac{G\,m_{\rm 1}\,m_{\rm 2}}{m_{*}\,\sigma^{2}},\\
 a_{\rm ej} = \frac{\xi{} \,{}m_3^2}{(m_1+m_2)^{3}}\,{}\frac{G \,{}m_1\,{} m_2}{v_{\rm esc}^{2}},\\
 a_{\rm gw}=\left[\frac{32\,{}G^2}{5\,{}\pi{}\,{}\xi{}\,{}c^5}\,{}\frac{\sigma{}\,{}m_1\,{}m_2\,{}(m_1+m_2)}{\rho{}_{\rm c}\,{}(1-e^2)^{7/2}}\,{}f_1(e)\right]^{1/5}.
\end{eqnarray}
Here, $ a_{\rm hard}$ is the limit for a binary to be considered hard \citep[][]{heggie1975}, $a_{\rm ej}$ is the maximum semi-major axis for ejection by three-body encounters, and $a_{\rm gw}$ is the limit below which the semi-major axis shrinking by the emission of GWs becomes dominant with respect to dynamical hardening. In the above equations, $m_{*}$ is the average mass of a star in the cluster, $\sigma$ the typical 3D velocity dispersion  of the cluster, $v_{\rm esc}{\sim {2\sigma}}$ the escape velocity, $\rho{}_{\rm c}$ the star cluster core density, and $\xi{\sim{3}}$ a numerically calibrated constant \citep[][]{Hills1993,Quinlan1996}. The mean of the log-normal distribution of the semi-major axes was computed as the average of the logarithmic limits, such that it resulted in $\mu_{\log{(a/{\rm AU})}}=2.42, 1.22, 0.42$ in the cases of YSCs, GCs, and NSCs, respectively. The dispersion was derived from the simulations of \cite{dicarlo2019} and \cite{dicarlo2020a}, and set as $\sigma_{\log{(a/{\rm AU})}}=0.92$ in all three samples of simulations, as was already done in \cite{dallamico2021}. With this prescription, all our BBHs are hard binaries for which GW emission is negligible if compared to hardening, but at the same time their semi-major axis is large enough such that previous interactions did not lead to a dynamical ejection of the binary from the cluster.

We assumed that the BHs were in thermal equilibrium with the rest of the cluster core so that the initial velocity at infinity, $v_{\infty}$, of the single BH with respect to the centre of mass of the binary could be sampled from a Maxwell-Boltzmann distribution \citep{heggie1975}. For the three sets of simulations, we assumed a 3D velocity dispersion of 5, 20, and 50~km~s$^{-1}$ in the cases of YSCs \citep[][]{zwart2010}, GCs \citep[]{prior1993}, and NSCs \citep{Neumayer2020}, respectively. 

We set the initial distance of the single BH from the centre of mass of the binary as $D=10^3\,{} a$. This guaranteed that the BBH had not been perturbed by the intruder before the beginning of the integration. 

The impact parameter, $b$, of the interaction was sampled from a uniform probability distribution proportional to $b^{2}$ \citep{hut1983} in the range $[0,b_{\rm max}]$, with $b_{\rm max}$ defined as
\begin{equation}\label{eq:phinney}
b_{\rm max}=\frac{\sqrt{\,{}2\,{}G\,{}(m_{1}+m_{2}+m_{3})\,{}a}}{v_{\infty}}.
\end{equation}
This is the maximum impact parameter for a strong three-body interaction with a hard binary derived by \cite{Sigurdsson1993}. 
 As $v_{\infty}$ and $a$ change between YSCs, GCs, and NSCs, 
the impact parameter in these three environments will also be different. 
Equation~\ref{eq:phinney} assumes strong gravitational focusing, that is, $G\,{}(m_1+m_2+m_3)/(v_\infty^2\,{}b)\gg{}1$, and a minimum intruder-binary star distance, $r_{\rm p}=a$. In Appendix~\ref{sec:bmax}, we discuss the impact of this assumption on our results.

LIGO--Virgo observations favour a population of low-spin BHs \citep{gwtc3}. Therefore, we extracted the magnitude of the dimensionless spin, $\chi$, of each BH from a Maxwell-Boltzmann distribution with root mean square $\sigma_\chi{}=0.1$ and truncated to $\chi{}=1$. Furthermore, we isotropically sampled the spin directions, accounting for the fact that dynamical encounters randomise them \citep{Rodriguez2016c}.


\section{Results}

\subsection{Outcomes}

\begin{figure*}
\begin{center}
	\includegraphics[width=1.6\columnwidth]{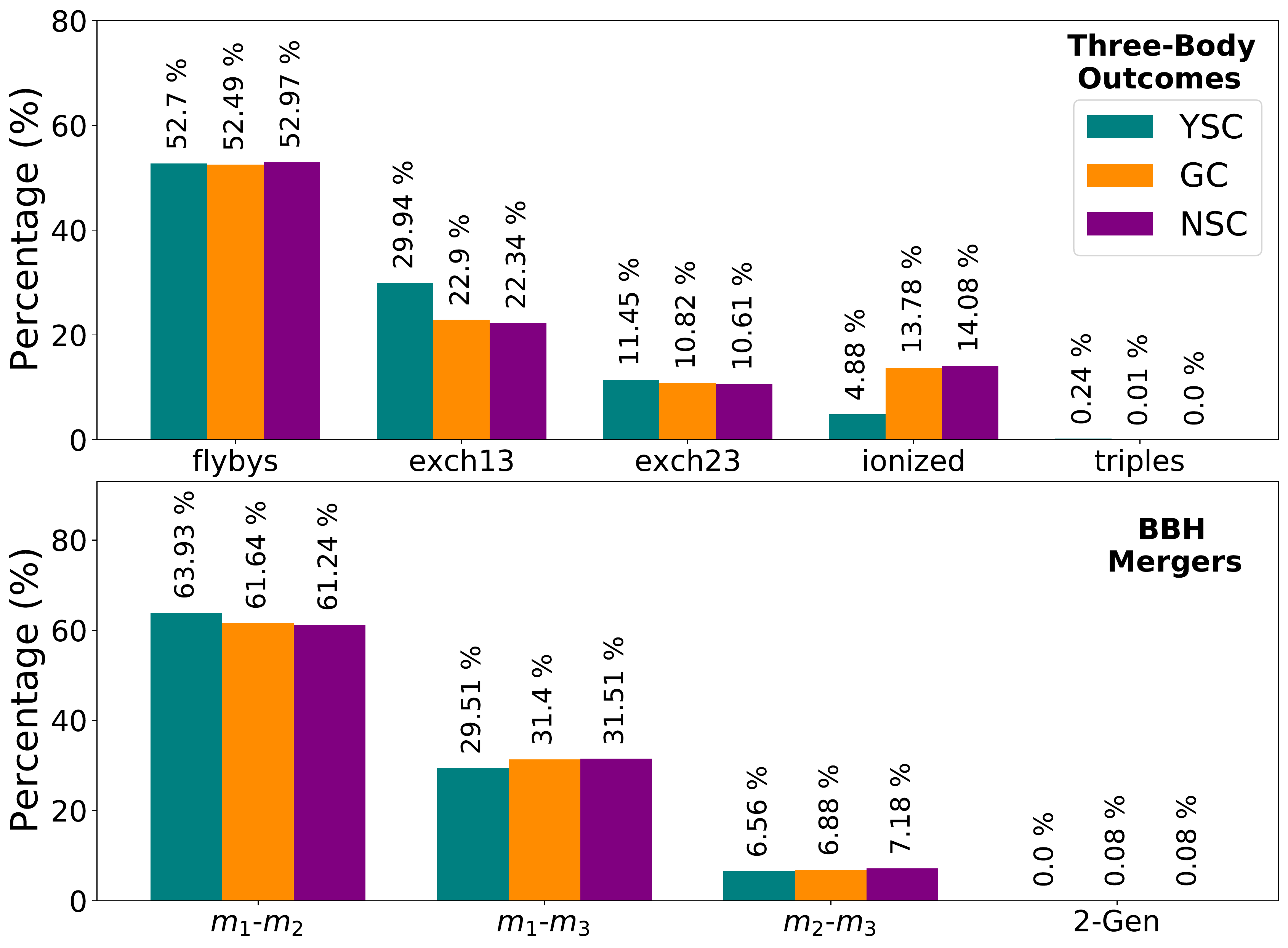}
    \caption{Upper panel: Percentage of different interaction outcomes 
    for each cluster type. From left to right: (i) flyby events, (ii) exchanges in which the intruder replaces the secondary BH, (iii) exchanges in which the intruder replaces the primary BH, (iv) ionisations, and (v) unstable triples. Lower panel: 
    Percentage of BBH mergers. From left to right:  (i)  BBH mergers occurring after a flyby, (ii) an exchange interaction with the secondary BH replaced by the intruder, (iii) an exchange with the primary BH replaced by the intruder, and (iv) second-generation BBH mergers. In both the upper and lower panels, the colours mark the cluster type in which the interaction takes place.}
    \label{fig:barplot}
\end{center}
\end{figure*}

 We divided the outcomes of the interactions into five classes, as a function of the system configuration at $1\,$Myr since the beginning of the simulation. We classified each simulation as follows.
 \begin{itemize}
     \item \textbf{Flyby:} the final binary has the same components as the initial binary. If the encounter hardens it enough, this binary may merge during or after the simulation. 
     
     \item \textbf{Exch13:} the three-body interaction ends with a binary composed of the primary BH of the initial binary and the intruder ($m_{\rm 1}-m_{\rm 3}$). If this exchanged binary merges during the simulation, we still classify the interaction as an exch13 event.

     \item \textbf{Exch23:}  the three-body interaction ends with a binary composed of the secondary BH of the initial binary and the intruder ($m_{\rm 2}-m_{\rm 3}$).  If this exchanged binary merges during the simulation, we still classify the interaction as an exch23 event.
     
     \item \textbf{Ionisation:} the encounter splits the initial binary, resulting in three single BHs.
     
     \item \textbf{Triple:} the system is still in an unstable triple configuration at the end of the simulation.
     
 \end{itemize}

The upper panel of Fig.~\ref{fig:barplot} classifies the end states of our binary-single scattering experiments. 
Flybys are the most frequent end state in all three cluster types, followed by exchanges. This result is expected, since these encounters generally have a larger impact parameter, $b$, than the semi-major axis of the initial binary, $a$: if $b\gg{}a$ the intruder sees the binary as a point-like object, and the interaction evolves into a flyby.

In YSCs, the number of ionisations is lower with respect to both GCs and NSCs. Vice versa, exchanges are more common in YSCs than in both GCs and NSCs. This happens because the typical dispersion velocities of YSCs are around $5\,$km$\,$s$^{-1}$, much smaller than GCs and NSCs, where the intruder can likely have a velocity higher than the critical velocity required to break up the binary system \citep{hut1983}. At $1\,$Myr, unstable triples  are much more numerous in YSCs  than in GCs, while we find no triple systems in NSCs at the end of our simulations. In YSCs, the encounter takes place later than in more dense clusters, since the interparticle distance is much larger while the dispersion velocity is lower. The intruder takes more time to reach the binary and the interaction begins at later times in the simulation, resulting in 
several systems that at $1\,$Myr are still in an unstable triple configuration.


\subsection{BBH mergers}

\begin{table}
	\begin{center}
	\caption{Percentage of peculiar events in YSCs, GCs, and NSCs.}
	\label{tab:merger}
	\begin{tabular}{lccc} 
		\hline
		     Event & P$_{(\%)}^{\rm YSC}$ & P$_{(\%)}^{\rm GCs}$ & P$_{(\%)}^{\rm NSC}$ \\
        \hline
        \hline
        
        Merger & $0.2$ & $2.4$ & $11.8$\\
        Merger $t_{\rm coal}<1\,{\rm Myr}$ & $14.8$ & $4.6$ & $4.0$\\ 
        PIBH & $44.3$ & $33.3$ & $27.7$ \\
        IMBH & $1.6$ & $0.7$ & $0.4$\\
        Merger with $e_{\rm 10\,Hz}>0.1$ & $1.6$ & $1$ & $0.6$\\
        Merger with $e_{\rm 1\,Hz}>0.1$ & $9$ & $1.8$ & $0.9$\\
        Retained BH remnants & $0$ & $3.1$ & $19.9$\\
		\hline
	\end{tabular}
	\end{center}
		\flushleft
\footnotesize{Line 1: Percentage of BBH mergers over all the three-body simulations. Line 2: Percentage of BBH mergers that take place during the three-body simulation with {\sc arwv} over all the BBH mergers. Line 3: Percentage of pair-instability BHs produced by the merger. Line 4: Percentage of intermediate-mass BHs (IMBHs) produced by the merger. Line 5: Percentage of BBH mergers with an eccentricity at a GW frequency of 10 Hz $e_{10\,\rm Hz}>0.1$ with respect to all BBH mergers. Line 6: Percentage of BBH mergers with an eccentricity at a GW frequency of 1 Hz $e_{1\,\rm Hz}>0.1$ with respect to all BBH mergers. Line 7: Percentage of BHs produced by BBH mergers that happen before star cluster evaporation and 
are retained inside the cluster after the three-body interaction and the GW recoil.}
\end{table}

We find that $0.2\%$, $2.4\%$, and $11.8\%$ of the simulations produce BBHs that merge within a Hubble time (13.8 Gyr) in YSCs, GCs, and NSCs, respectively. Of these mergers, $14.8\%$, $4.6\%$, and $4\%$ take place in the first Myr of integration with {\sc arwv}.  
These results, which are also reported in Table~\ref{tab:merger}, imply that the denser and more massive the cluster is, the larger the fraction of mergers produced via three-body interactions. This happens because the minimum binding energy of a hard binary is higher in more massive clusters. As a result, BBHs in NSCs are consistently closer to the GW regime than in YSCs, making it more likely for a single interaction to push them into the GW emission regime. The lower panel of Fig.~\ref{fig:barplot} classifies these mergers by their formation channel: flyby, exch13, or exch23, and second-generation mergers (i.e., mergers that occur between the remnant of a previous merger and the third BH). 
The bar plot shows that flybys account for $\sim53\%$ of all our encounters, but $\sim61-64\%$ of the BBH mergers. Hence, flybys are more efficient in inducing mergers than, for example, exch23 events, which in turn are $\sim 11\%$ of the outcomes but account only for $\sim 7\%$ of the mergers. This happens because exchanges usually produce new BBHs with a larger total mass but also with a larger semi-major axis than BBHs involved in flyby events.  

\begin{figure}
\begin{center}
	\includegraphics[width=1.\columnwidth]{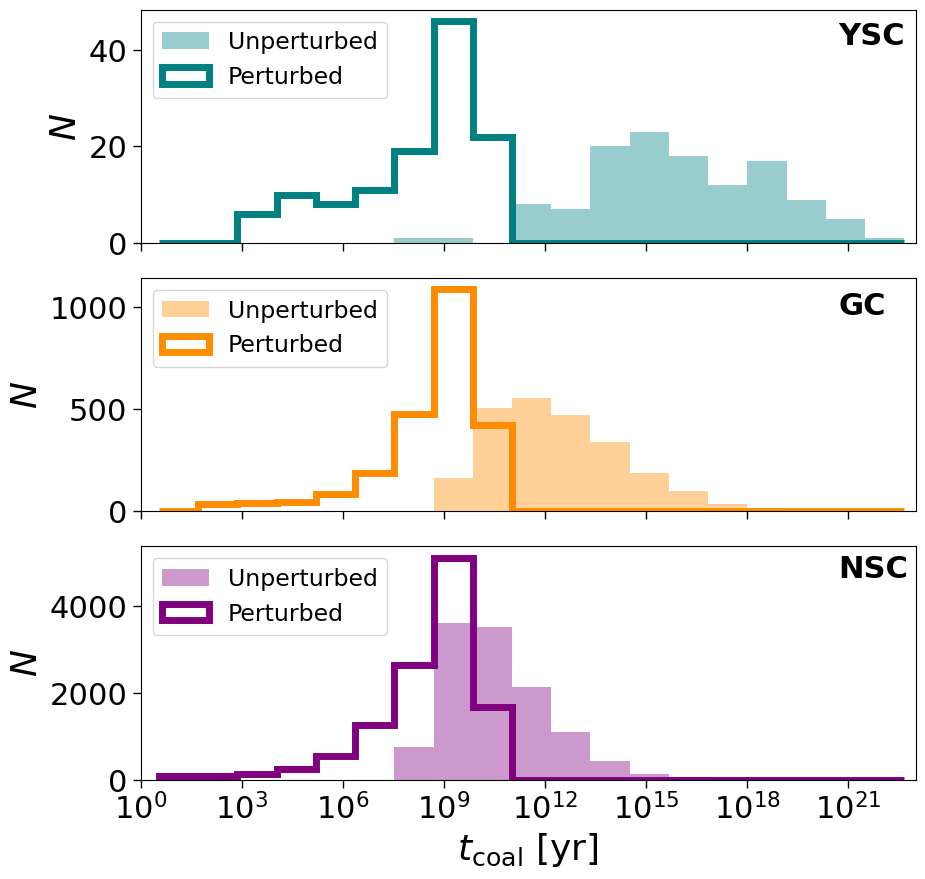}
    \caption{Distribution of coalescence times for unperturbed initial binaries (filled histograms), and for the same initial binaries perturbed by the three-body interaction (unfilled histograms). The upper, central, and lower panels show the cases of YSCs, GCs, and NSCs,  respectively.}
    \label{fig:tcoal}
\end{center}
\end{figure}

For some systems that produce a BBH merger, we find that the initial binary would have merged within a Hubble time even without the interaction. These are binaries that, if evolved as an unperturbed binary, would have produced a BBH merger without undergoing any three-body encounter. These systems are $1.6\%$, $11.4\%$, and $45.0\%$ of the BBH mergers in YSCs, GCs, and NSCs respectively. Even for these systems, the interaction has efficiently sped up the merger. This can be seen in Fig.~\ref{fig:tcoal}, which shows the coalescence time of the unperturbed binaries (if the three-body interaction would not have taken place), and the coalescence time resulting from the three-body simulation. 
For instance, all the BBH mergers that take place during the simulation with {\sc arwv} within the first $1\,$Myr rapidly merge solely as a consequence of the three-body interaction.

Finally, in a few cases, we find the formation of second-generation BBH mergers. In these simulations, two BHs merge during the three-body interaction, and their remnant forms a new BBH with the remaining BH, which in turn is able to reach coalescence in less than a Hubble time.  
We find the same percentage  of second-generation mergers in GCs and NSCs, while no second-generation systems form in YSCs via three-body interactions (Fig.~\ref{fig:barplot}).

\begin{figure}
	\includegraphics[width=1.\columnwidth]{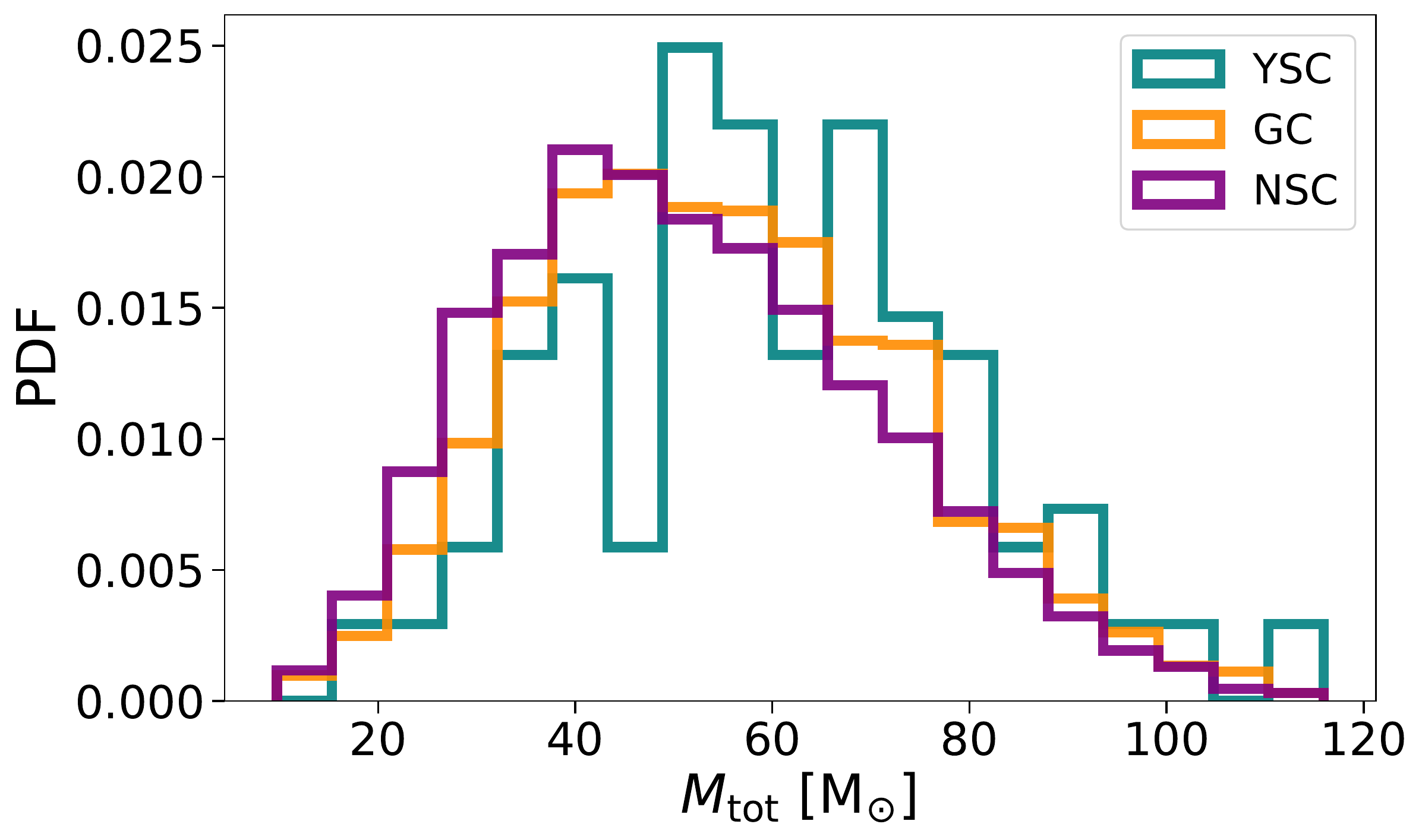}
    \caption{Total mass of the BBH mergers in YSCs (blue), GCs (orange), and NSCs (purple).}
    \label{fig:mtot}
\end{figure}

Despite the three simulation sets being initialised with the same BH mass spectrum, the total mass of the BBH mergers produced in YSCs, GCs, and NSCs differ, as is shown by Fig.~\ref{fig:mtot}. This difference in the mass of BBH mergers is clearly an effect of the environment.  

Mergers by dynamical exchange are slightly favoured in GCs and NSCs with respect to YSCs (Fig.~\ref{fig:barplot}). Since exchanges typically take place when the intruder is more massive than at least one of the two binary components, we should expect more massive BBH mergers in GCs and NSCs. Nevertheless, Fig.~\ref{fig:mtot} shows that YSCs are more likely to produce massive mergers via three-body encounters than the other two sets of simulations. This happens because in YSCs only the most massive systems are able to merge within a Hubble time.
In GCs and NSCs, given the larger velocity dispersions, the hard-soft boundary is shifted towards the lower semi-major axis (see eq.~\ref{eq:smalimits}). Therefore, GW emission can also be efficient for relatively low-mass BBHs (eq.~\ref{eq:peters}). In YSCs, instead, given the larger semi-major axis at formation, mergers are more biased towards the most massive BBHs. In this way, NSCs and GCs are more efficient in the production of BBH mergers with low-mass components than YSCs. This is further confirmed if we look at the percentage of massive BH remnants produced by these mergers in Table~\ref{tab:merger}: pair-instability BHs, defined as BHs with mass in the range $60-100\,$M$_{\odot}$,  are $44.3\%$, $33.3\%$, and $27.7\%$ over all the mergers produced in YSCs, GCs, and NSCs, respectively. Intermediate-mass BHs (IMBHs, i.e. BHs with a mass $\geq{}10^2$ M$_\odot$) are  $1.6\%$, $0.7\%$, and $0.4\%$ over all the mergers in YSCs, GCs, and NSCs, respectively. The most massive remnant produced by a merger is $\sim110\,$M$_{\odot}$ in all three sets. In YSCs and GCs, the most massive remnants are produced by a flyby event, while in the NSC case the most massive remnant is produced by the merger of a second-generation BBH. 

Pair-instability BHs and IMBHs form mainly via flybys, because these are the most frequent type of interaction among our simulations. 
Exch13 are the events most likely to create a BBH more massive than the initial binary. On the other hand, 
exch23 tend to produce less massive binaries than the initial ones. Nevertheless, the latter have a minor impact on the total mass distribution in Fig.~\ref{fig:mtot} because they are rare events, representing only $\approx{7}$\% of all mergers (Fig.~\ref{fig:barplot}). Finally, mergers of second-generation binaries are rare even among pair-instability BHs and IMBHs: they represent $0.25\%$ and $0\%$ of pair-instability BHs and IMBHs in GCs, and $0.12\%$ and $4\%$ in NSCs, respectively. 

\subsection{Eccentric mergers}\label{sec:ecc}

\begin{figure*}
    \centering
	   \includegraphics[width=\linewidth]{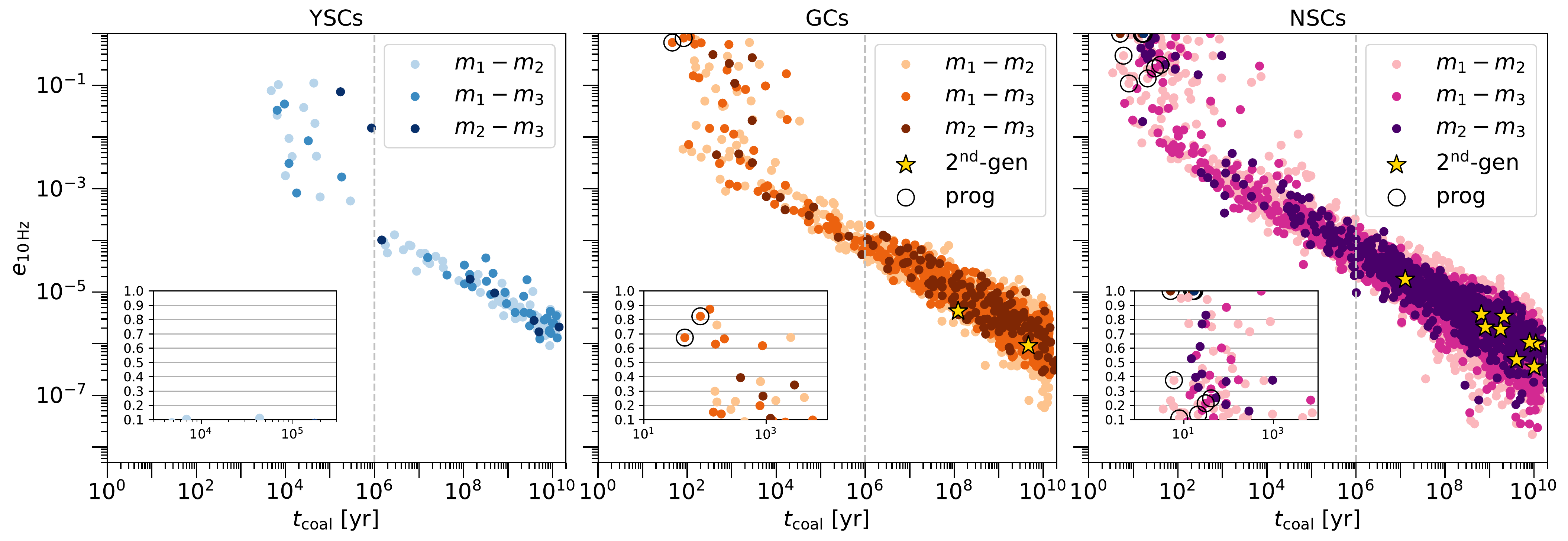}
        \caption{Orbital eccentricity at 10 Hz ($e_{\rm 10\,{}Hz}$)  as a function of the coalescence time ($t_{\rm coal}$) of BBH mergers in YSCs (left), GCs (centre), and NSCs (right).
        Different colours indicate mergers through flybys and the two families of exchanges. The yellow stars show second-generation BBH mergers. The empty black dots highlight the  BBH progenitors of second-generation mergers. The dashed vertical line divides the plot into two regions: up to $1\,$Myr we evolve the systems with {\sc arwv}, above $1\,$Myr all the dynamical interactions are concluded, and with Eqs.~\ref{eq:peters} we stop the {\sc arwv} integration and evolve the remaining BBHs. The three insets show a zoom for mergers with $e_{\rm 10\,{}Hz}>0.1$. We note that some of the points at $e_{\rm 10\,{}Hz}\sim1$ overlap.}
    \label{fig:ecc}
\end{figure*}

Figure~\ref{fig:ecc} shows the eccentricity of the BBH when the emitted GW frequency is 10 Hz in the source frame (hereafter, $e_{\rm 10\,{}Hz}$) as a function of the coalescence time from the beginning of the simulation. All three sets present two distinct families of mergers regarding their eccentricity. In the first one, and also the most common one, the mergers follow the relation, $t_{\rm coal}(e)$, of eq.~\ref{eq:peters}, such that $e_{\rm 10\,{}Hz}$ 
decreases as the coalescence time of the binary increases. In these systems, the dynamical encounter ends before the merger takes place. This allows the BBH resulting from the interaction to evolve unperturbed up to the merger, such that the eccentricity evolution is ruled only by the angular momentum loss by GW emission.\footnote{After 1 Myr all the mergers are forced to follow eq.~\ref{eq:peters}, since we assumed no further dynamical interactions after the first encounter. This choice is discussed more in detail in Section~\ref{sec:sing3}.}

The second family of mergers is composed by systems that do not follow the relation of eq.~\ref{eq:peters}, but rather reach the merger with 
$e_{\rm 10\,{}Hz}>10^{-3}$ in a relatively short time during the simulation with {\sc arwv}, while the dynamical interaction is still ongoing. Some of these systems even have $e_{\rm 10\,{}Hz}>0.1$ (insets of Fig.~\ref{fig:ecc}).  In the following, we refer to the mergers with $e_{\rm 10\,{}Hz}>0.1$ simply as eccentric mergers. These are $1.6\%$, $1\%$, and $0.6\%$ of all the mergers in YSCs, GCs, and NSCs, respectively (Table~\ref{tab:merger}). 
Although YSCs exhibit a larger fraction of eccentric mergers, the only two eccentric events occurring in YSCs both have $e_{\rm 10\,{}Hz}\sim{0.1}$. Conversely, eccentric events in GCs and NSCs span eccentricities up to and beyond $e_{\rm 10\,{}Hz}=0.9$. We can divide these eccentric mergers as a function of the type of interaction that triggered the coalescence:

\begin{itemize}
    \item \textbf{Chaotic mergers} are the product of 
    three-body interactions in which temporary binaries with brand-new orbital parameters are continuously formed and destroyed. If the eccentricity of these temporary binaries is sufficiently high, and their lifetime is longer than the perturbation timescale of the outer body, a nearly radial merger can be triggered. These are the most common interactions to produce eccentric mergers, accounting for $\sim{63}\,\%$ of all the eccentric events with $e_{\rm 10\,{}Hz}>0.1$ in GCs and NSCs. The left-hand panel of Fig.~\ref{fig:traj} shows one of these interactions and the subsequent eccentric merger.

    \item \textbf{Prompt mergers} are the second most common event to cause eccentric mergers, representing $\sim{31}\,\%$ of all these events in GCs and NSCs. The right-hand panel of Fig.~\ref{fig:traj} shows an example of an eccentric merger that follows a prompt interaction. These mergers typically follow flyby events in which the intruder significantly extracts angular momentum from the binary, driving the two components into a nearly radial orbit and inducing a prompt merger. This is for example the case of the two eccentric mergers in the YSC set. We also call prompt mergers simulations in which the intruder tangentially intersects the orbital plane of the binary and approaches one of the two components with an almost anti-parallel velocity, such that they rapidly merge in a nearly head-on collision (as the simulation in the right-hand panel of Fig.~\ref{fig:traj} demonstrates). This is the case of the most eccentric mergers in our simulations. In NSCs, five head-on collisions trigger a merger, with $e_{\rm 10\,{}Hz}\sim1$ (right-hand panel of  Fig.~\ref{fig:ecc}), while in GCs the maximum value of $e_{\rm 10\,{}Hz}$ is 0.87 (central panel Fig.~\ref{fig:ecc}). 

    \item \textbf{Temporary triple mergers} take place when the system evolves as a hierarchical triple. Stable triple systems cannot form from three isolated 
    unbound bodies \citep[][]{Naoz2016}; however, temporary stable hierarchical triples can be created via three-body interactions. In this configuration, the intruder sets in an outer orbit, perturbing the initial binary and causing it to merge rapidly enough that GW emission does not efficiently circularise the binary's orbit. This is the case for the system shown in the central panel of Fig.~\ref{fig:traj}. The merger occurs only if the system remains stable for a sufficient period of time for the perturbations to be effective.
    Binary systems involved in chaotic mergers exhibit considerably shorter lifespans than those within temporary triple systems. In contrast, temporary triple systems can survive for several orbital periods of the third body, which has enough time to orbit around the inner binary perturbing it. We define an eccentric merger as a temporary triple merger if the merger event involves the inner binary and if the outer orbiting third body is able to carry out at least two complete stable orbits with approximately the same period around the inner binary. If these conditions are not met, we define the merger as a chaotic merger.
    Due to the low stability of these systems, temporary triple mergers are responsible only for $\sim6\,\%$ of the eccentric mergers in GCs and NSCs.
    
    
\end{itemize}

Even if flybys are the most common formation path of BBH mergers in our simulations (lower panel of Fig.~\ref{fig:barplot}), eccentric mergers come in almost the same proportion from flybys and exchanges 
in both GCs and NSCs. Hence, eccentric mergers are more likely to form from exchanges. No second-generation BBH mergers belong to eccentric mergers. With our assumption of no further dynamical interactions beyond 1$\,$Myr, second-generation binaries, after their formation, circularise their orbits before reaching coalescence. On the other hand, all the mergers that give birth to one of the two BHs that compose second-generation binaries are eccentric mergers. We refer to these systems as BBH progenitors of second-generation mergers.
In the GC scenario, the progenitor systems of the two second-generation BBHs are both eccentric mergers with an eccentricity of $0.82$ and $0.67$, respectively, with the first being an almost head-on collision in a prompt event, and the latter coming from a chaotic merger. This is also true for the NSC case, in which all the nine progenitors have $e_{\rm 10\,{}Hz}>0.1$, four of which merge in a head-on collision with $e_{\rm 10\,{}Hz}\sim{1}$.

\begin{figure*}
    \centering
	   \includegraphics[width=\linewidth]{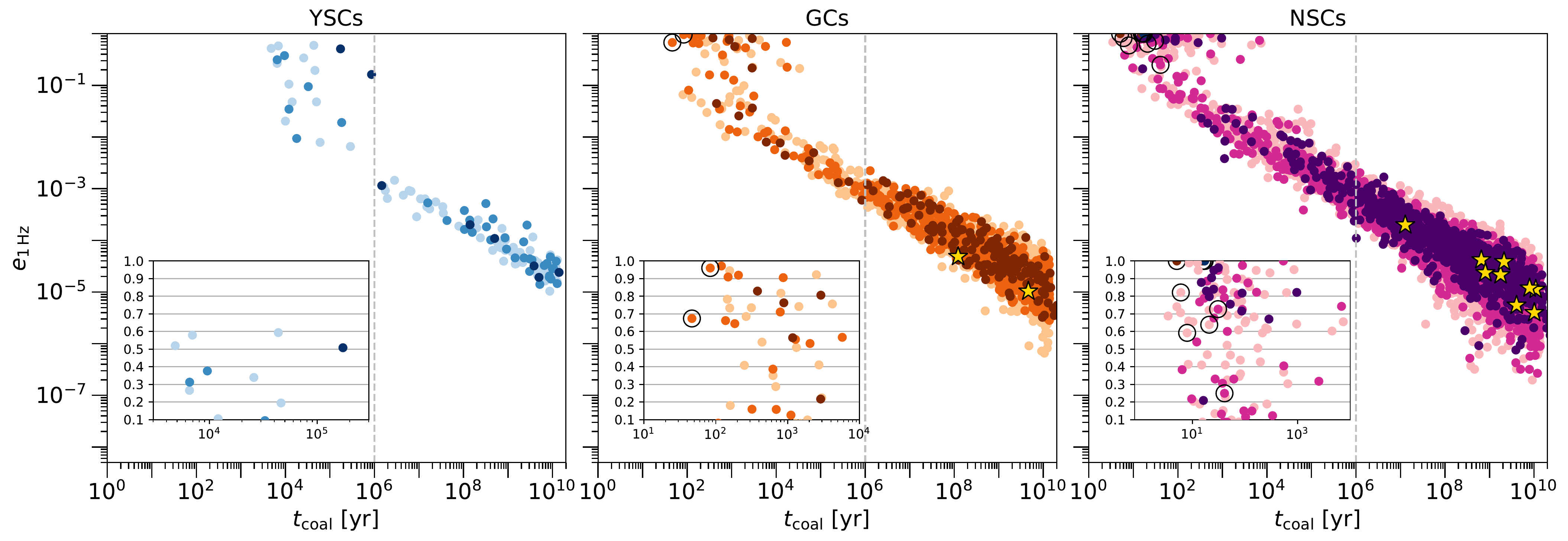}
        \caption{Same as fig.~\ref{fig:ecc} but with the orbital eccentricity computed at 1 Hz ($e_{\rm 1\,{}Hz}$).}
    \label{fig:eccET}
\end{figure*}

Finally, Fig.~\ref{fig:eccET} shows the same systems as Fig.~\ref{fig:ecc} but with the eccentricity computed at the GW frequency of 1 Hz in the source frame. This frequency is the predicted lower limit in the sensitivity band of the Einstein Telescope \citep{Punturo2010,Maggiore2020}. In this regime, eccentric mergers constitute $0.9\%$, $1.8\%$, and $9\%$ of all mergers in NSCs, GCs, and  YSCs, respectively (Table~\ref{tab:merger}). This is because, at lower frequencies, the two BHs are at a larger orbital separation and GWs still have to efficiently circularise the orbit, such that more systems have higher eccentricities.

\subsection{Dynamical ejections}

Our BBHs can escape from the cluster 
as a consequence of dynamical recoil. 
Three-body interactions with hard binaries tend to reduce the binary internal energy and convert it into kinetic energy of the system. This induces a recoil velocity, $v_{\rm rec}$, both on the binary and the single object. If this velocity is larger than the escape velocity, $v_{\rm esc}$, of the cluster, both the binary and the single object are dynamically ejected from the cluster \citep[][]{heggie1975,Hills1975,Sigurdsson1993}. 

\begin{figure}
     \centering
     \begin{subfigure}[b]{0.5\textwidth}
         \centering
         \includegraphics[width=\textwidth]{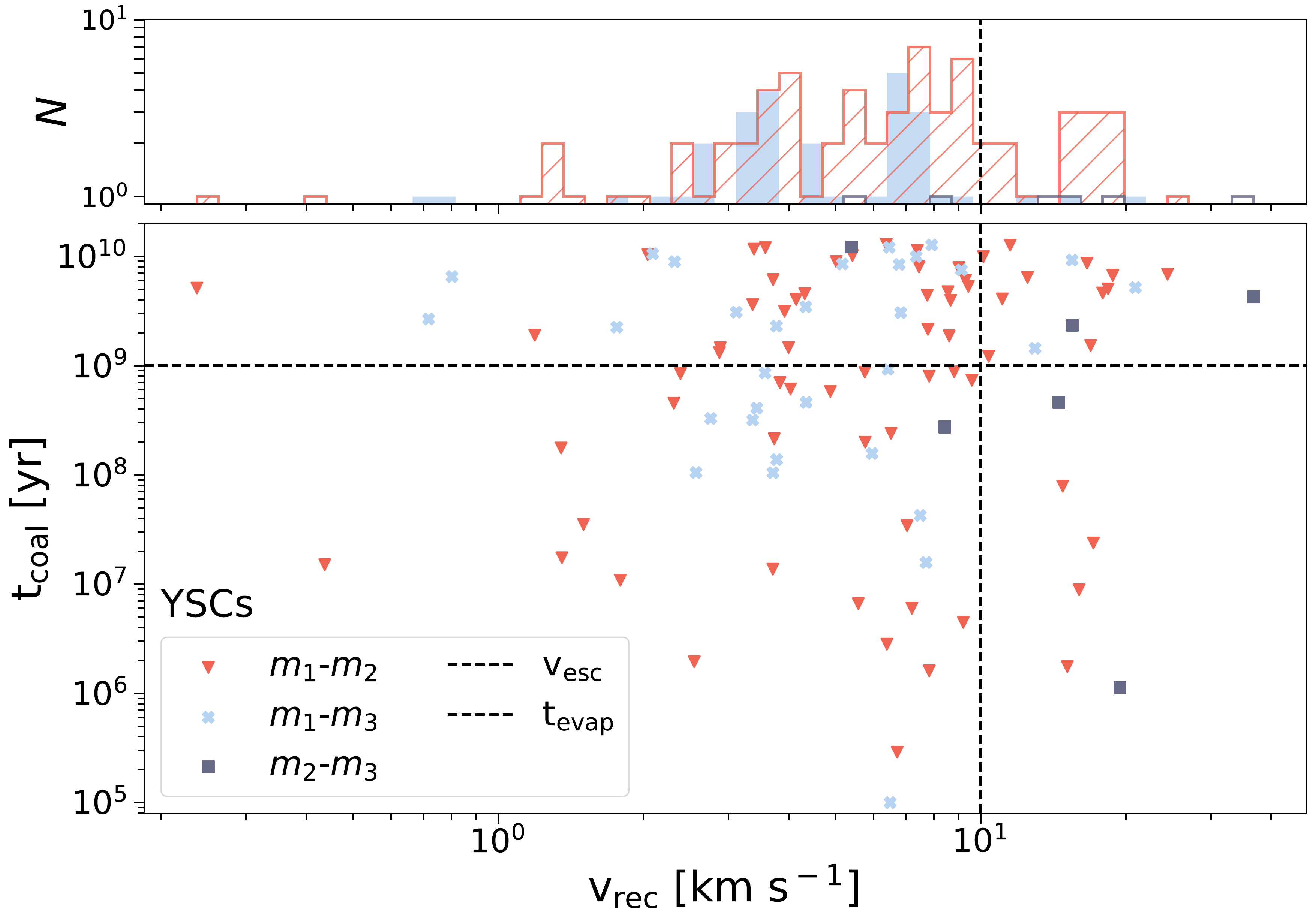}
         \label{fig:esc_ysc}
     \end{subfigure}
     \hfill
     \begin{subfigure}[b]{0.5\textwidth}
         \centering
         \includegraphics[width=\textwidth]{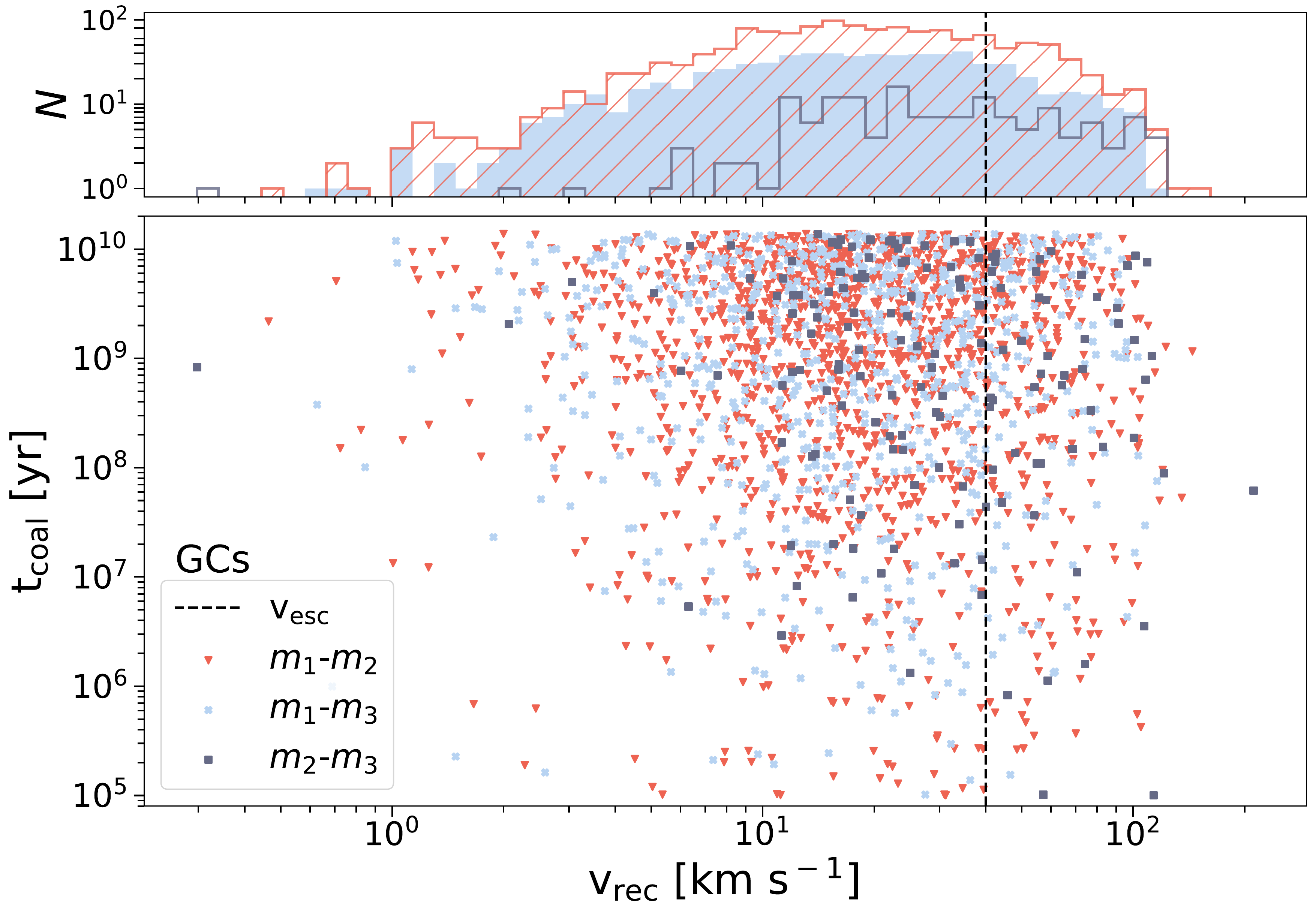}
         \label{fig:esc_gc}
     \end{subfigure}
     \hfill
     \begin{subfigure}[b]{0.5\textwidth}
         \centering
         \includegraphics[width=\textwidth]{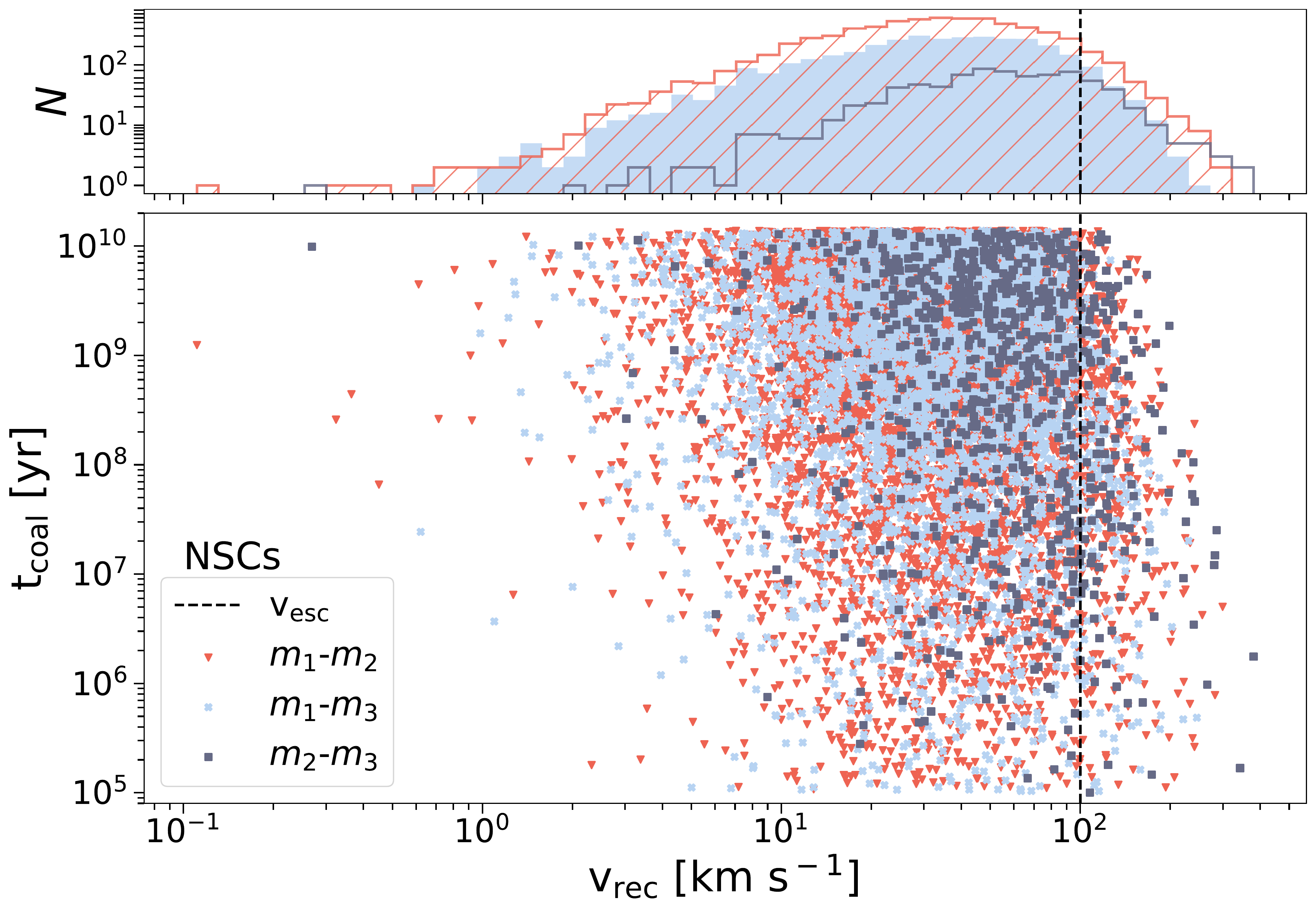}
         \label{fig:esc_nsc}
     \end{subfigure}
        \caption{Upper, mid, and lower scatter plots display the BBH mergers that take place after the three-body interactions in YSCs, GCs, and NSCs, respectively. All the mergers that take place during the three-body interaction, i.e. $<10^{6}\,$yr, are assumed to merge inside the cluster. The x axis reports the recoil velocity caused by the energy exchange in the three-body encounter. The y axis shows the coalescence time of the binary from the beginning of the simulation. Different outcomes are shown with different markers and colours. In the YSCs case, the horizontal dashed line reports the typical evaporation time of a YSC. The vertical dashed line in all the plots shows the escape velocity of the cluster. 
        Distributions of the recoil velocities are displayed as marginal histograms. The colour legend is the same as the scatter plot.}
        \label{fig:escapers}
\end{figure}

This has major implications for BBH mergers: on the one hand, the energy exchange speeds the merger up, reducing the semi-major axis and thus the coalescence time, t$_{\rm coal}$; on the other hand, it might also kick out the binary from the cluster, preventing further encounters. If the ejection happens before the binary merges, the BH remnant produced by the coalescence will not be able to dynamically interact with other bodies of the cluster and produce new binaries. This may strongly affect the efficiency of the hierarchical merger mechanism by which repeated BH mergers produce massive BHs \citep[e.g.][]{miller2002,mapelli2021a}.
Figure~\ref{fig:escapers} shows the impact of these dynamical ejections on BBH mergers. 
Here, we assume that all the mergers that take place during the three-body encounter merge inside the cluster. Therefore, the plots report only the BBH mergers that take place after $10^{5}\,$yr and for which the three-body interaction is concluded. These recoil velocities span from less than $1\,$km$\,$s$^{-1}$ up to $\sim40\,$km$\,$s$^{-1}$ in YSCs, $\sim200\,$km$\,$s$^{-1}$ in GCs, and $\sim400\,$km$\,$s$^{-1}$ in NSCs. These differences are, once again, explained by the large binding energy of the binaries in GCs and NSCs, which translates into larger recoil kicks. In YSCs, for example, the BBHs have a larger semi-major axis than binaries in more dense clusters and the recoil velocities are lower.

In the three simulation sets, the most likely population of ejected mergers is the one produced by exch23 events.
The marginal plots of Fig.~\ref{fig:escapers} and Table~\ref{tab:rec} show that these BBHs generally have larger recoil velocities with respect to the other two families of mergers. The fraction of BBH mergers ejected after the three-body interaction 
is $6\%$ in NSCs, $18\%$ in GCs, and $18\%$ in YSCs. In YSCs, we must also keep into account the evaporation of the cluster; that is, when the cluster dissolves because of stellar mass loss and tidal stripping
\citep[][]{Spitzer1987,Heggie2003,Binney2008}. Due to evaporation, most of the BBH mergers in YSCs 
happen in the field even without being dynamically ejected \citep[e.g.][]{rastello2021,torniamenti2022}. If we assume a typical evaporation time, t$_{\rm evap}\sim1\,$Gyr for a YSC\footnote{This evaporation timescale refers to YSCs with mass $\sim10^{4}\,$M$_\odot$ and must be considered as an upper limit. Processes like galactic perturbations and encounters with giant molecular clouds might, in principle, accelerate the disruption of the cluster \citep[e.g.][]{Gieles2006}. } \citep[][]{torniamenti2022},  $48\%$ of the BBH mergers happen inside the cluster and $5\%$ take place in the field because of dynamical recoil, while the remaining $47\%$ occur in the field because of cluster evaporation. Finally, all the eccentric mergers in GCs and NSCs merge inside the cluster. They are mostly the product of chaotic interactions that take place over a short timescale and rapidly lead to the merger of two of the three BHs.

\begin{figure}
	\includegraphics[width=1.\columnwidth]{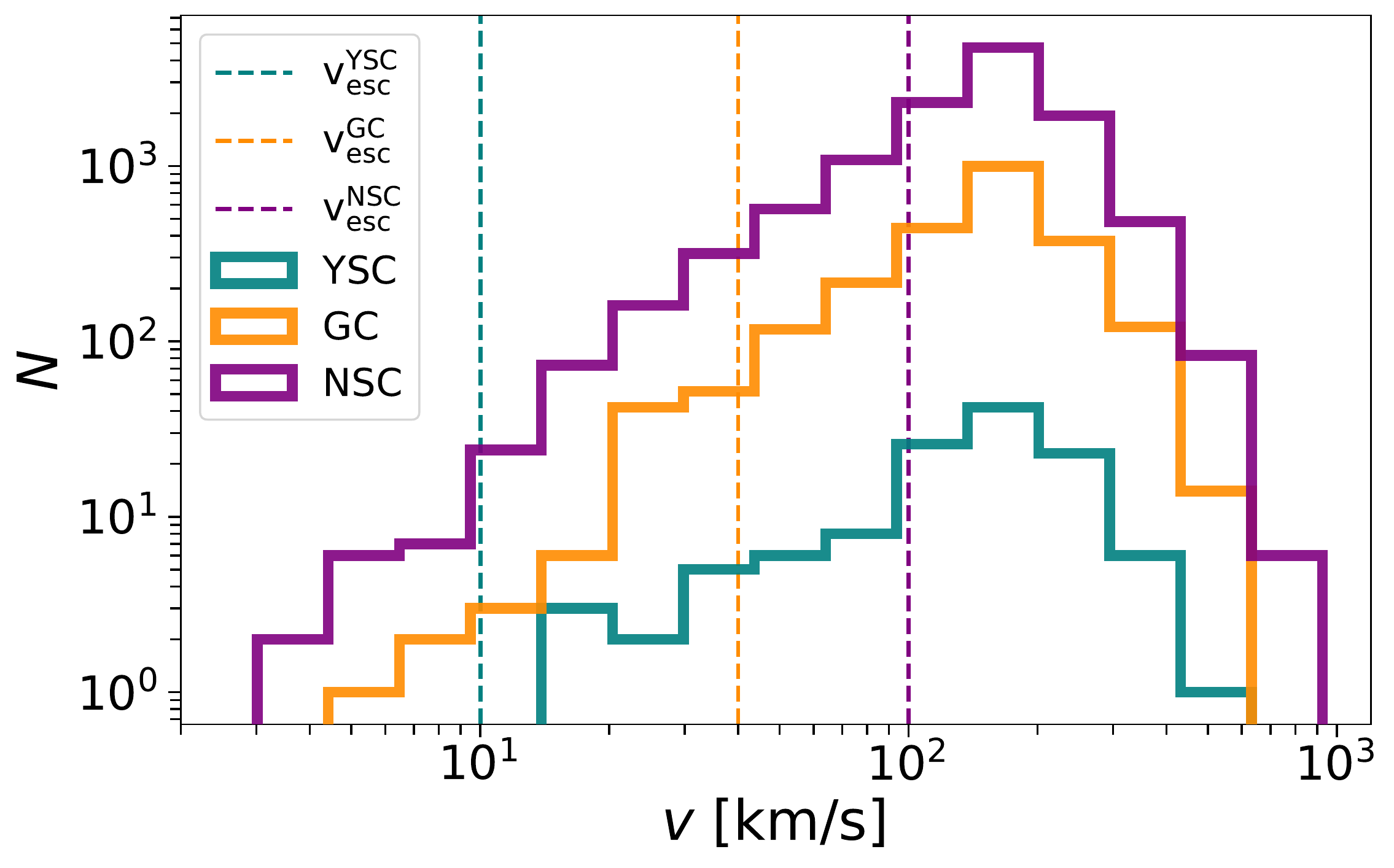}
    \caption{Distribution of the relativistic recoil kick of the remnants produced by the BBH mergers in YSCs, GCs, and NSCs. The dashed vertical lines represent the escape velocities of the three types of star clusters.}
    \label{fig:vkick}
\end{figure}

GWs emitted from a spiraling BBH are generally irradiated anisotropically due to the asymmetry of the system. This induces linear momentum transfer on the remnant BH produced by the merger, which translates into a relativistic kick that might accelerate the remnant even up to a few thousand of km$\,$s$^{-1}$ \citep[]{Fitchett1983,Maggiore2018}. Figure~\ref{fig:vkick} shows this relativistic kick produced by all the BBH mergers in our three sets of simulations. 
All three distributions peak at $\sim200\,$km$\,$s$^{-1}$, with velocities that are approximately one order of magnitude higher than the ones reported in Fig.~\ref{fig:escapers}.
Hence, GW recoils are more efficient in ejecting BHs from their parent cluster than three-body recoils. Due to GW kicks, only 
$3.6\%$ and $21.2\%$ 
 of the BH remnants are retained in GCs and NSCs, respectively. This fraction drops to $0$ in YSCs.

We can now count the overall fraction of BH remnants retained by the cluster for which 1) the relativistic recoil is below the escape velocity of the cluster, 2) the BBH progenitor is not ejected after a three-body interaction, and 3) the binary is able to merge inside 
the cluster before its evaporation. This fraction is $0\%$ for YSCs, $3.1\%$ for GCs, and $19.9\%$ for NSCs (Table~\ref{tab:merger}). 
Finally, all the second-generation BH remnants in GCs and NSCs are kicked out of the cluster due to GW recoils.

\begin{table}
	\begin{center}
	\caption{Three-body recoil velocities.}
	\label{tab:rec}
	\begin{tabular}{lccc} 
		\hline
		     Cluster & $v_{\rm rec}^{m_{\rm 1}-m_{\rm 2}}$ & $v_{\rm rec}^{m_{\rm 1}-m_{\rm 3}}$ & $v_{\rm rec}^{m_{\rm 2}-m_{\rm 3}}$ \\
                    & [km$\,$s$^{-1}$]  & [km$\,$s$^{-1}$]   &  [km$\,$s$^{-1}$]  \\
        \hline
        \hline
        
        YSC & $7$ & $4$ & $15$\\
        GC & $18$ & $18$ & $26$\\
        NSC & $33$ & $34$ & $55$\\
        
		\hline
	\end{tabular}
	\end{center}
		\flushleft
\footnotesize{Median recoil kicks of three-body interactions in the YSC, GC, and NSC cases. The values refer to the distributions of the marginal plots in Fig.~\ref{fig:escapers} and are reported for the three different types of BH couples that merge after a three-body encounter.}
\end{table}

\subsection{Orbital plane tilt}

\begin{figure}
	\includegraphics[width=1.\columnwidth]{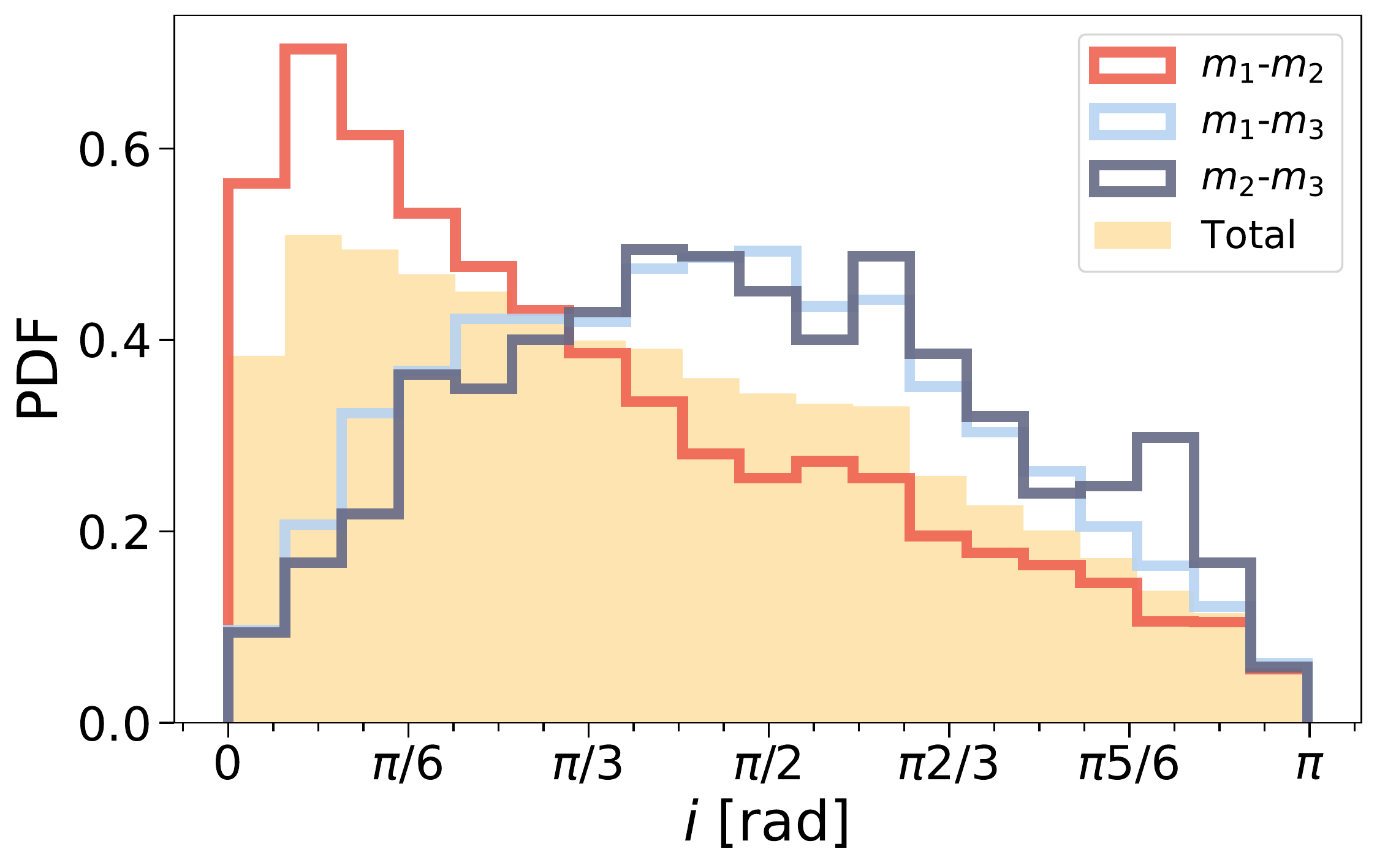}
    \caption{Tilt angle distributions of the BBH orbital plane in NSCs at the end of the simulation with respect to the initial binary orbital plane. Unfilled histograms are flybys (red), exch13 (blue), and exch23 (grey). The filled histogram (yellow) shows the total distribution. We do not show GCs and YSCs because they have exactly the same behaviour.}
    \label{fig:tilt}
\end{figure}

Figure~\ref{fig:tilt} shows the tilt angle, $i$, defined as the angle between the orbital plane of the initial binary and the orbital plane of the final binary left after the three-body interaction. These distributions show that three-body interactions are capable of inducing large tilt angles in the BBH population, in good agreement with what was found by \citealt{Trani2021} \cite[see also][]{banerjee2023}. The magnitude of the tilt depends on the outcome of the interaction. Binaries that underwent a flyby generally experience smaller tilt angles with respect to exchanges, with the distribution peaking at $\sim15^{\circ}$ for the former and $\sim90^{\circ}$ for the latter. Since there is no strong correlation between the initial angle, $\theta$ (sec.~\ref{sec:IC}), of the interaction and the outcome, the distributions of Fig.~\ref{fig:tilt} are a direct product of the interactions. This means that flybys statistically induce small perturbations in the orbital plane of the initial binary if compared to exchanges. Exchanges, on the other hand, favour the production of more massive newborn binaries with an orbital plane likely tilted with respect to the original one. This implies that the orientation of spins in dynamically assembled BBHs might not be perfectly isotropic.
Our finding is consistent with \cite{bouffanais2019}, who assumes an isotropic spin orientation for exchanges, but nearly aligned spins for flybys, which are less perturbed by the encounter.

\section{Discussion and caveats}\label{sec:disc}

\subsection{Merger rate density of eccentric mergers}
In Section~\ref{sec:ecc}, we derived the fraction of eccentric mergers in YSCs, GCs, and NSCs; that is, those that have $e_{\rm 10\,{}{\rm Hz}}>0.1$. 
Through semi-analytical simulations, \cite{Mapelli2022b} were able to derive the merger rate density of BBH mergers in YSCs, GCs, and NSCs. They find $\mathcal{R}_{\rm YSC}=3.0\,{}$Gpc$^{-3}\,{}$yr$^{-1}$, $\mathcal{R}_{\rm GC}=4.4\,{}$Gpc$^{-3}\,$yr$^{-1}$, and $\mathcal{R}_{\rm NSC}=1.3\,$Gpc$^{-3}\,$yr$^{-1}$ in their fiducial model. These rates refer to first-generation BBH mergers. 
Since our first-generation BHs 
are drawn from the same population as \cite{Mapelli2022b}, we combined these rates with the fraction of eccentric mergers produced by three-body interactions in our three simulation sets. In this way, we find the following rates of eccentric mergers as a function of the type of cluster: $\sim{2\times{}10^{-3}}\,$Gpc$^{-3}\,$yr$^{-1}$ in YSCs, $\sim{4\times{}10^{-2}}\,$Gpc$^{-3}\,$yr$^{-1}$ in GCs, and $\sim{8\times{}10^{-3}}\,$Gpc$^{-3}\,$yr$^{-1}$ in NSCs. These rates represent a rough estimate, since a more realistic evaluation requires taking into account the possible differences between the coalescence times of our eccentric mergers and the BBH merger population of \cite{Mapelli2022b}.

\subsection{Single versus multiple three-body interactions}\label{sec:sing3}
Our eccentric mergers (Fig.~\ref{fig:ecc}) qualitatively match the properties of eccentric mergers simulated by 
\cite{Samsing2017b} and 
\cite{Zevin2019}. For example, \cite{Samsing2017b} and \cite{samsing2018b} find that $\sim{1\%}$ of all the mergers by binary-single encounters in GCs are eccentric mergers, which is in excellent agreement with the fraction of eccentric events that we find in this work. On the other hand, \cite{samsing2018a} and \cite{Rodriguez2018} find a fraction of eccentric mergers $\sim{5}$ times higher than the one we find in our simulations. This difference is a consequence of our choice to simulate one single interaction for each binary.

Assuming a single dynamical encounter per binary is particularly well founded in YSCs, since three-body interactions between BHs are expected to happen with a frequency of $\sim1$ per cluster \citep{dicarlo2019}. In more dense clusters, it is more likely for a BBH to experience multiple interactions during its lifetime. However, as we presented in Section~\ref{sec:ecc} and in Fig.~\ref{fig:ecc}, all the eccentric mergers take place while the dynamical interaction is still ongoing, meaning that in our simulations the typical timescale for an eccentric three-body merger is smaller compared to the timescale of a three-body interaction. Our results imply that even one single three-body encounter in a BBH lifetime is sufficient to speed up the merger and to produce an eccentric event in the LIGO--Virgo band. Nevertheless, the masses, eccentricities, and fraction of ejected BHs reported in this work must be interpreted as lower limits, since further interactions might produce more massive and eccentric mergers and more dynamical ejections. This is especially true for all the mergers that take place after $1\,$Myr in our simulations. The effects of multiple three-body interactions on the BBH population are going to be presented in a future paper.


Finally, our work does not consider the special case of an active galactic nucleus (AGN) disc. In AGN discs, eccentric mergers can be significantly boosted if binary-single scatters occur with small mutual inclinations (of less than a few degrees). This ultimately leads to a relatively flat distribution of the spin-tilt angle in eccentric mergers \citep{Samsing2022}, which is very different from the one we find here for NSCs. The reason for this difference is that three-body interactions in star clusters require completely distinct initial conditions with respect to the AGN case. We will include AGN discs in forthcoming studies.


\section{Summary}

In this work, we have presented the results of $2.5\times10^{5}$ three-body simulations performed via direct \textit{N}-body integration with the {\sc arwv} code \citep{mikkola1989,chassonnery2019}. Our simulations incorporate post-Newtonian corrections up to the 2.5th order and 
adopt initial conditions that mimic the properties of YSCs, GCs, and NSCs. With this approach, we aim to investigate the influence of the host environment on: 1) the outcomes of three-body encounters, 2) the populations of BBH mergers produced through interactions, and 3) the production of BBH mergers with non-negligible eccentricities in the LIGO--Virgo frequency range. Our results can be summarised as follows.

\begin{itemize}

    \item We divide the outcomes into flybys, exchanges in which the primary or the secondary BH component is replaced by the intruder, ionisations, and triples. Flybys dominate the interactions in all the simulation sets, accounting for approximately $\sim53\%$ of all the outcomes. YSCs differ from GCs and NSCs, with fewer ionisations (around $\sim5\%$ compared to $\sim14\%$ for GCs and NSCs) but more exchanges (about $\sim41\%$ compared to $\sim33\%$ for GCs and NSCs), and also a non-zero number of systems that are in an unstable triple configuration at the end of the simulation. 

    \item Three-body interactions in GCs and NSCs produce a higher number of mergers compared to YSCs. Approximately $2.4\%$ and $11.8\%$ of the simulations in GCs and NSCs, respectively, lead to a BBH merger within a Hubble time, compared to the $0.2\%$ of the simulations in YSCs. Flybys are the most effective pathway to produce mergers as they significantly decrease the coalescence time. 
    Of the three types of clusters we considered, YSCs are less efficient at producing low-mass BBH mergers than both GCs and NSCs. 
    
    \item Pair-instability BH remnants ($60-100\,$M$_{\odot}$) are $\sim44\%$, $\sim33\%$, and $\sim28\%$ of all the mergers produced in YSCs, GCs, and NSCs, while IMBHs ($>100\,$M$_{\odot}$) are $1.6\%$, $0.7\%$, and $0.4\%$, respectively. Finally, we find second-generation BBH mergers only in GCs and NSCs, accounting for $0.08\%$ of all mergers in both sets.
    
    \item The percentage of BBH mergers with an orbital eccentricity higher than $0.1$ at a GW frequency of $10\,$Hz in the source frame ($e_{\rm 10\,{}Hz}>0.1$)
    is $1.6\%$, $1\%$, and $0.6\%$ in YSCs, GCs, and NSCs, respectively. 
    We additionally investigated the percentage of BBH mergers with an eccentricity exceeding $0.1$ at a GW frequency of $1\,$Hz in the source frame. This frequency corresponds to the lower boundary of the sensitivity band of the Einstein Telescope. We find that $9\%$ of all the BBH mergers in YSCs are eccentric in this frequency range, while this fraction drops to $1.8\%$ for mergers in GCs and $0.9\%$ for mergers in NSCs. 

    \item In both GCs and NSCs, the most frequent interactions leading to eccentric mergers are chaotic exchange events, accounting for approximately $63\%$ of all eccentric mergers. These involve the creation and destruction of several temporary binaries before the merger takes place. Prompt interactions, including flybys in which the intruder extracts enough angular momentum from the binary to cause a radial merger, and head-on collisions between the intruder and one of the binary components, account for approximately $31\%$ of eccentric events. Finally, BBH mergers in temporary stable hierarchical triples contribute to approximately $6\%$ of all eccentric mergers in GCs and NSCs. In our simulations, all the progenitors of second-generation BBHs are eccentric mergers in both GCs and NSCs.

    \item The percentage of remnant BHs that are not expelled from the cluster is $0\%$ for YSCs, $3.1\%$ for GCs, and $19.9\%$ for NSCs. These are BHs that are not dynamically ejected from the cluster by the three-body and GW relativistic recoil kicks, and for which the progenitor BBHs merge before the evaporation of the star cluster. 
    In YSCs, $\sim50\%$ of the BBH mergers take place in the field after the cluster has evaporated. Relativistic recoil kicks due to anisotropic GW emission are the primary cause of dynamical ejections, with typical velocities that exceed $100\,$km$\,$s$^{-1}$. This strongly affects hierarchical mergers. 
    
    \item Three-body interactions alter the inclination of the original orbital plane, causing tilt angles that are not isotropically distributed, but that rather depend on the interaction outcome. 
    Exchanges tend to produce new binary systems that have an isotropically oriented orbital plane with respect to the original one, while flybys usually result in relatively minor perturbations of $\sim15^{\circ}$ on the orbital plane. This result challenges the idea that dynamics produces perfectly isotropic spin orientations.

    \item We estimate the merger rate density of eccentric BBH mergers to be $\sim{2\times{}10^{-3}}\,$Gpc$^{-3}\,$yr$^{-1}$ for YSCs, $\sim{4\times{}10^{-2}}\,$Gpc$^{-3}\,$yr$^{-1}$ for GCs, and $\sim{8\times{}10^{-3}}\,$Gpc$^{-3}\,$yr$^{-1}$ for NSCs. These rates must be regarded as lower limits, as we only considered a single three-body interaction per binary in our simulations. Additional dynamical interactions during the lifetime of these binaries may lead to an increase in the number of eccentric mergers.
    
\end{itemize}


\begin{acknowledgements}
      We are grateful to Roberto Capuzzo Dolcetta, Pauline Chassonnery, and Seppo Mikkola for making the {\sc arwv} code available to us. We thank Alessandro Trani for the useful comments. We also thank the members of the DEMOBLACK team for the helpful discussions. MD acknowledges financial support from Cariparo Foundation under grant 55440. MM and ST acknowledge financial support from the European Research Council for the ERC Consolidator grant DEMOBLACK under contract no. 770017. MAS acknowledges funding from the European Union’s Horizon 2020 research and innovation programme under the Marie Skłodowska-Curie grant agreement No. 101025436 (project GRACE-BH, PI: Manuel Arca Sedda).
\end{acknowledgements}

%
   \bibliographystyle{aa} 
   \bibliography{bibliography.bib} 


\begin{appendix}
\section{Impact of the maximum impact parameter}\label{sec:bmax}
Three-body scatterings are computationally advantageous with respect to full star cluster simulations. The price to pay is that  
the results depend on the choice of initial parameters, which must convey information about the properties of the host star clusters and their binary systems. 
For example, to compute the upper limit of the impact parameter, $b_{\rm max}$ (eq.~\ref{eq:phinney}), we assumed that the distance of closest approach, $r_{\rm p}$, is equal to the semi-major axis, $a$, of the binary system and that the strong gravitational focusing limit is valid, or in other words that $G\,{}(m_1+m_2+m_3)/(v_\infty^2\,{}b)\gg{}1$. These assumptions ensure that the majority of our three-body interactions are classified as hard encounters, where significant energy exchange occurs between the single and binary BHs. How critical is this assumption, and what are the effects on our results if we change this prescription? If we relax this hypothesis, the upper limit of the distribution takes the more general form
\begin{equation}\label{eq:otherb}
{b_{\rm max}^2 = r_{\rm p}^2\,{}\left[1 + 2\,{}\frac{G\,{}(m_1+m_2+m_3)}{r_{\rm p}\,{}v_\infty^2}\right].}
\end{equation}
 The distance of closest approach is typically defined as $r_{\rm p}=k\,{}a$, with $k$ assumed as an arbitrary constant. To study how the strong gravitational focus approximation  impacts our results, we ran $5\times10^{4}$ additional three-body interactions initialised as the main NSC set, but with impact parameters generated according to eq.~\ref{eq:otherb}, with $k=1,2,3,4,5$.

\begin{figure}
	\includegraphics[width=.9\columnwidth]{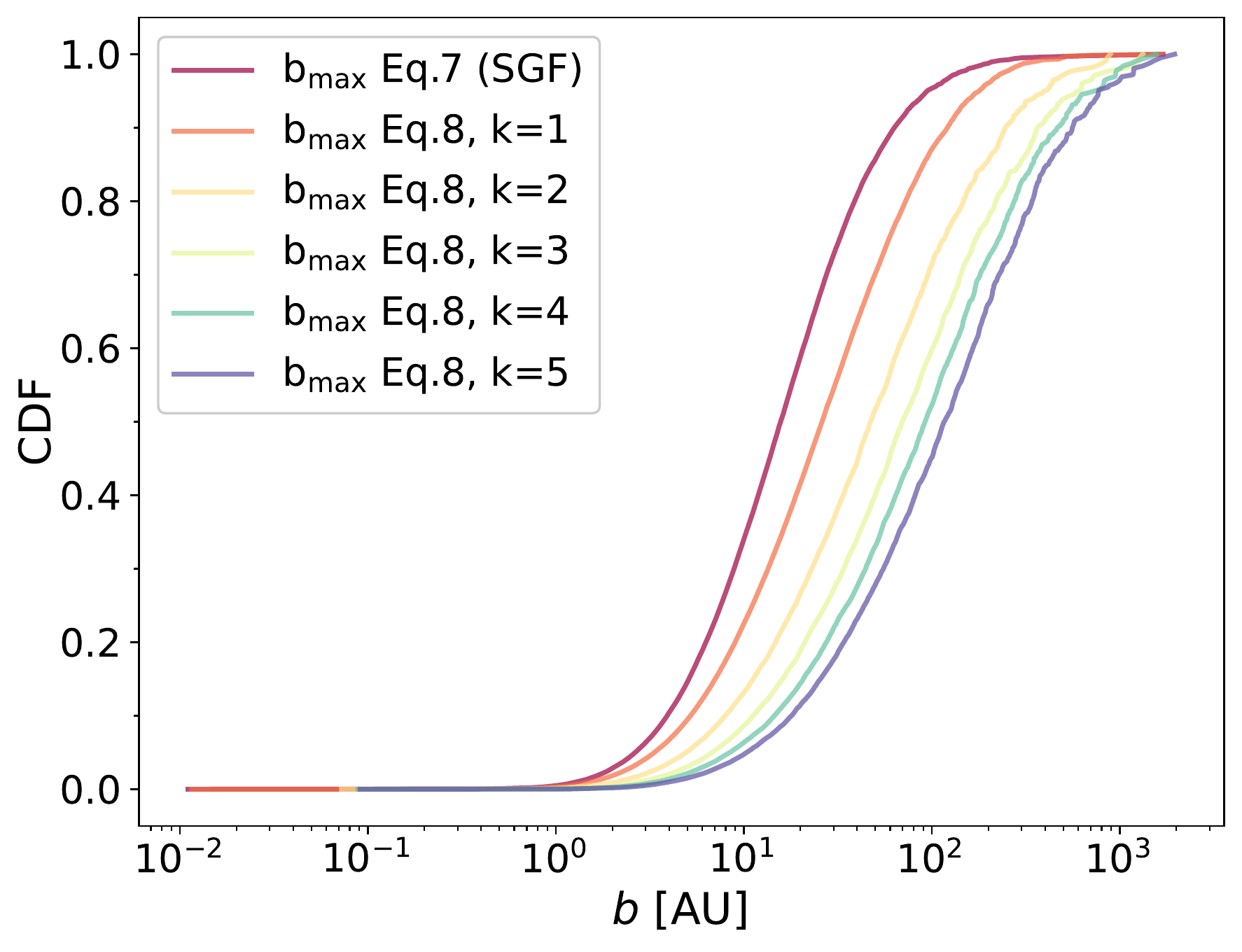}
    \caption{Cumulative distribution function of the impact parameter computed assuming b$_{\rm max}$ from eq.~\ref{eq:phinney}, and from eq.~\ref{eq:otherb} with k=1,2,3,4,5.}
    \label{fig:cumulativ}
\end{figure}

Figure~\ref{fig:cumulativ} shows 
the cumulative distribution function of the impact parameter 
if we consider $b_{\rm max}$ computed as in eq.~\ref{eq:phinney} or eq.~\ref{eq:otherb}, with various choices for $k$ in the NSC case. By calculating  $b_{\rm max}$  from eq.~\ref{eq:phinney}, it is more likely to generate interactions where the impact parameter favours closer encounters compared to the sample generated with eq.~\ref{eq:otherb}. The difference between the distributions is marginal with $k=1$, but it gradually grows with increasing $k$. For example, there is a probability of approximately $34\%$ for a simulation that has an impact parameter $\le10\,$AU in our fiducial model, while this probability decreases to $23\%$ for $k=1$ and drops to $5\%$ for $k=5$. 

The probability 
of drawing an impact parameter, $b>10^{3}\,$AU is $<0.1\%$ in our fiducial model, while this fraction rises to $3\%$ and $4\%$, respectively, in the cases of $k=4$ and $k=5$ with eq.~\ref{eq:otherb}. 
Assuming large values of $k$ produces a non-negligible fraction of simulations where the impact parameter is of the same order of magnitude as the interparticle distance in the core of a NSC. In these simulations, the single BH is likely to interact with other members of the star cluster before reaching the target binary, and the simulated interaction may not occur.

\begin{figure*}
\centering
	\includegraphics[width=1.5\columnwidth]{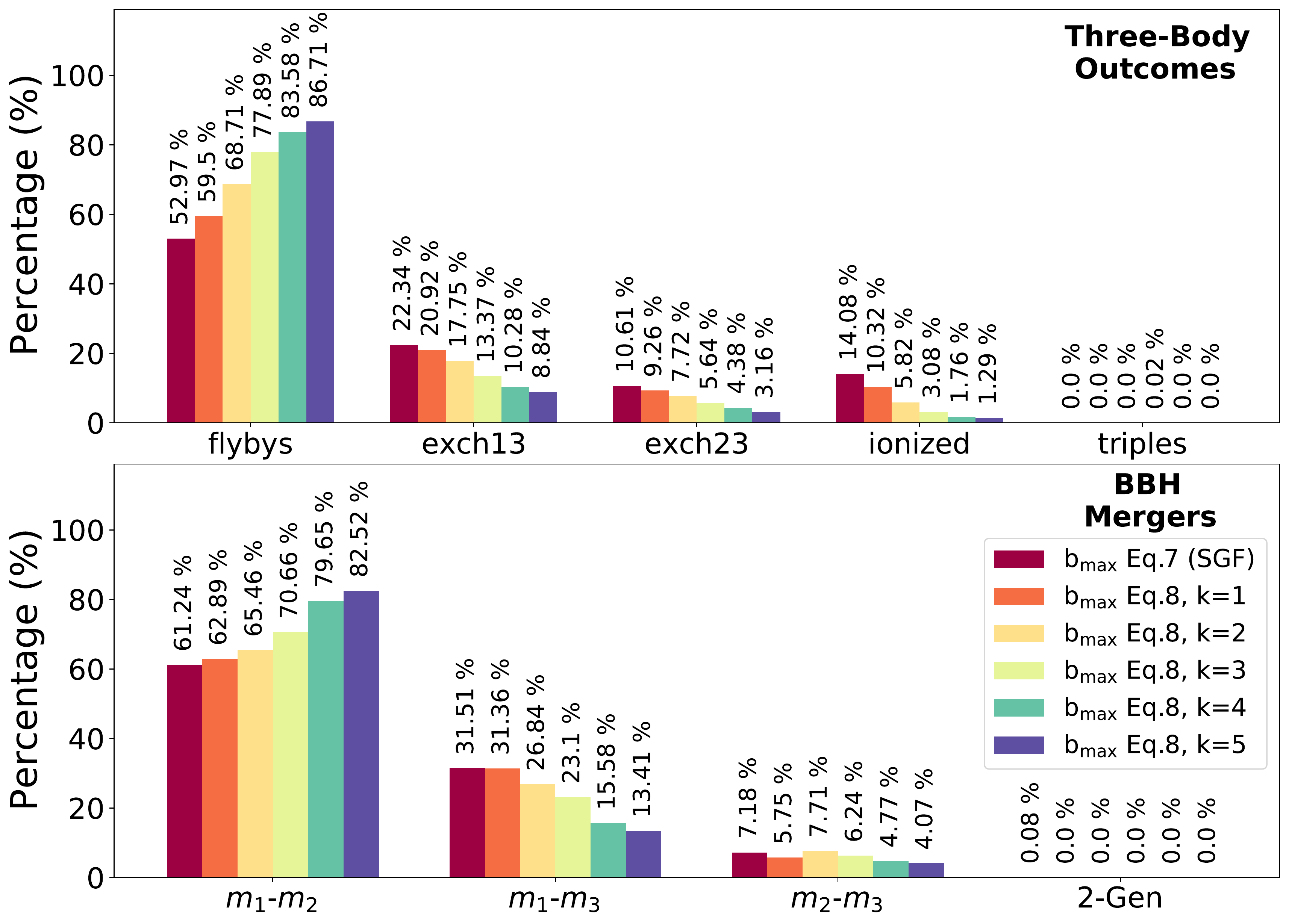}
    \caption{Same as Fig.~\ref{fig:barplot} but assuming eq.~\ref{eq:otherb} for the choice of $b_{\rm max}$. The plot shows the outcome of the three-body simulations (upper panel) and the formation channel of the BBH mergers (lower panel) for the main NSC set initialised with eq.~\ref{eq:phinney} (dark red, fiducial model) and the smaller NSC sets initialised with eq.~\ref{eq:otherb} assuming $k=1$ (orange), 2 (yellow), 3 (light green), 4 (dark green), and 5 (blue).}
    \label{fig:bmax}
\end{figure*}

Figure~\ref{fig:bmax} shows the outcomes of three-body interactions and  BBH mergers as a function of the assumed $b_{\rm max}$. The simulation sets with larger values of $b_{\rm max}$ result in fewer exchanges and ionisations while favouring more flybys, with this effect becoming more pronounced with larger values of $k$. For instance, in the cases of $k=4$ and $k=5$, almost eight to nine simulations out of ten evolve as a flyby. In contrast, our fiducial model produces flybys in almost 50~\% of cases. This happens because, when the strong gravitational focus approximation is not assumed, interactions with larger impact parameters become more common. 
In such encounters, the single BH is more likely to perceive the binary system as a point-like object, leading more frequently to weaker energy exchanges between the binary and the single object. It also disfavours exchanges, which require closer passages between the single BH and one of the two binary components.

\begin{figure}
	\includegraphics[width=1.\columnwidth]{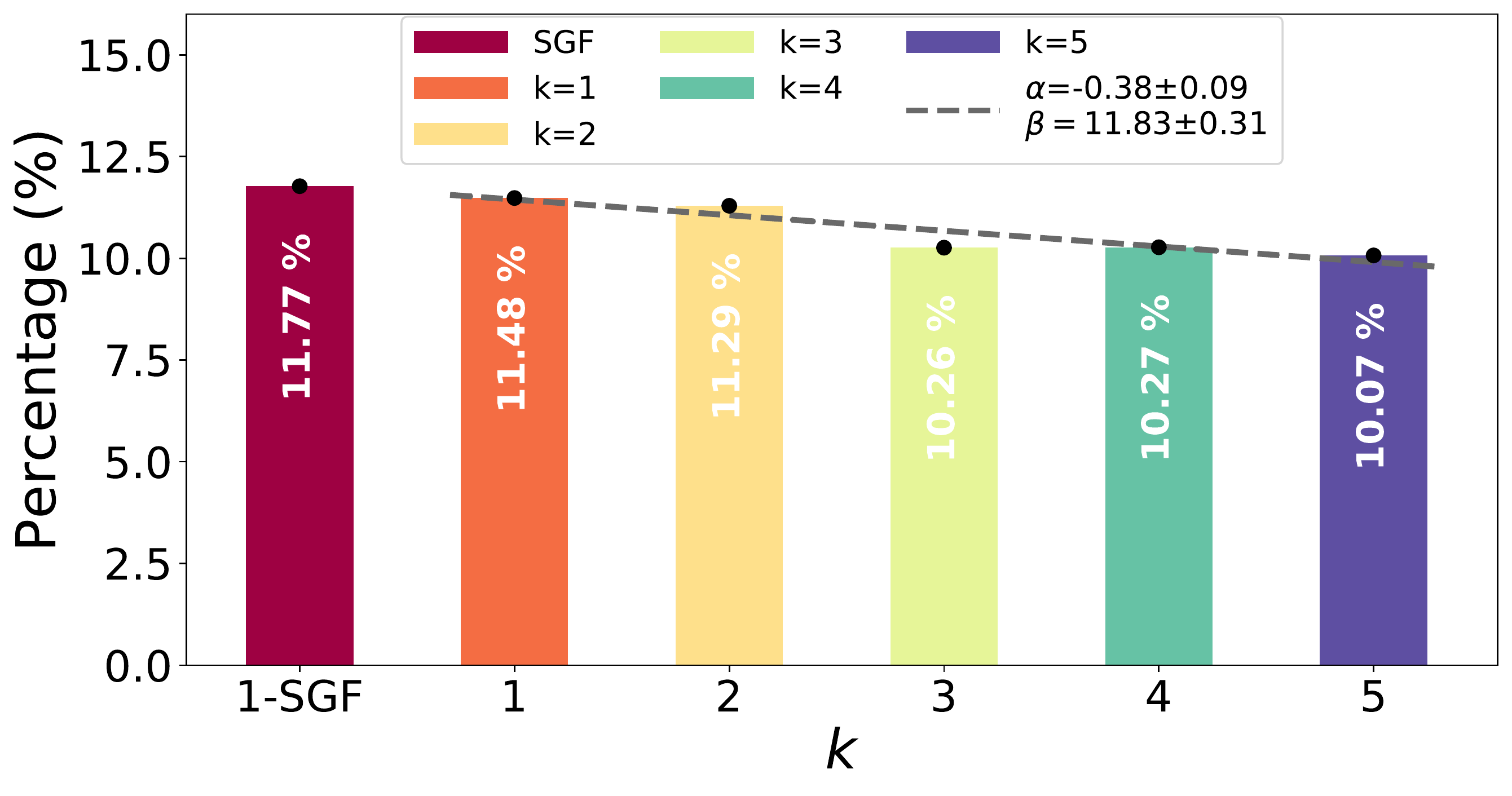}
    \caption{Percentage of BBH mergers over all the simulated encounters in NSCs,  where $b_{\rm max}$ was computed with eq.\ref{eq:phinney} and eq.\ref{eq:otherb}, assuming $k=1,\,{}2,\,{}3,\,{}4\,{},\,{}5$. The bar colours are the same as in Figs.~\ref{fig:cumulativ} and \ref{fig:bmax}. The dashed grey line is the linear fit to the percentage of mergers as a function of the $k$ parameter.}
    \label{fig:Pmerg}
\end{figure}

The lower panel of Fig.~\ref{fig:bmax} shows that the choice of $b_{\rm max}$ also affects the percentage of mergers, even if less dramatically than the percentage of flybys, exchanges, and ionisations.  
The dependence of the percentage of BBH mergers on the choice of $b_{\rm max}$ is further shown in Fig.~\ref{fig:Pmerg}. 
The fraction of BBH mergers decreases only by $1.7\%$ from our fiducial model to the simulation set with $k=5$. This plot also shows that the merger fraction linearly decreases with $k$ as $P_{\rm merg}=\alpha{}\,{}k+\beta$, with $\alpha=-0.38\pm0.09$ and $\beta=11.83\pm0.31$.


BBH mergers are less affected by the choice of $b_{\rm max}$ 
because both Eqs.~\ref{eq:phinney} and~\ref{eq:otherb} encode the distribution of our semi-major axis, which shapes the overall impact parameter distribution through the $b_{\rm max}$ dependence on $a$. In our approach, we sample $a$ in a way that ensures that all our binaries are hard, while also being sufficiently large to prevent efficient GW emission. 

Different approaches are assumed in the literature to sample the parameter space of close encounters. For example, \cite{hut1983} and \cite{Zevin2019} computed $b_{\rm max}$ as eq.~\ref{eq:otherb} and set $k=1$ with arbitrary fixed values for $a$. \cite{samsing2014} chose $k=5$ to ensure the sampling of distant interactions, but at the same time they kept the semi-major axis fixed at $10^{-5}\,$AU, which corresponds to very hard binaries, close to the merger. \cite{Quinlan1996}, on the other hand, fixed $a$ but varied the $k$ parameter.

Here, we adopted a physically motivated semi-major axis distribution, ranging from the limit between hard and soft binaries down to the threshold for efficient GW decay. Moreover, in our models we aim to fully explore the parameter space for hard encounters. The price that we pay is that we do not consider encounters with large $k$. Their effect is shown in fig.~\ref{fig:bmax} and \ref{fig:Pmerg}, for completeness.

\begin{figure}
	\includegraphics[width=1.\columnwidth]{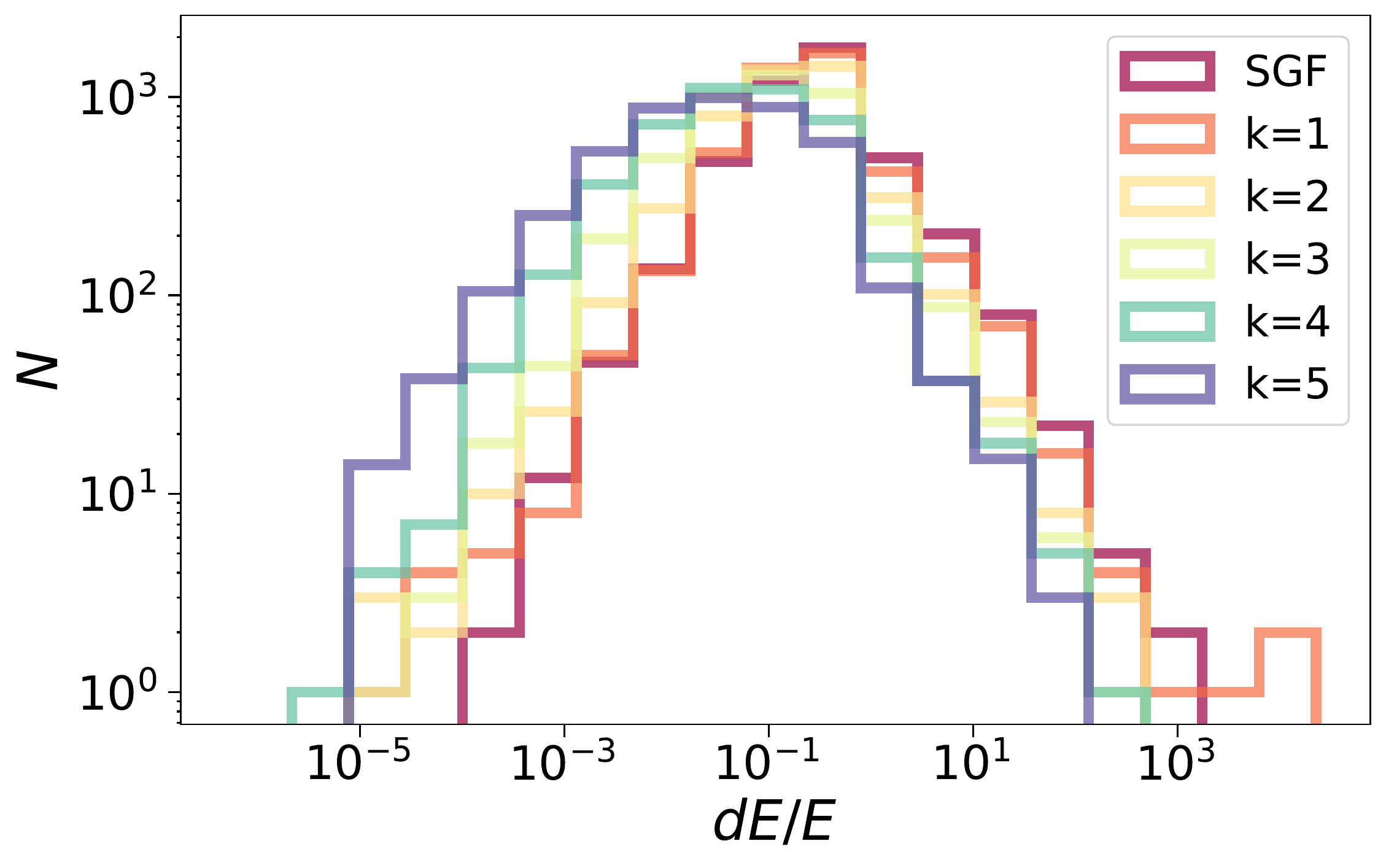}
    \caption{Histograms of the normalised variation in the binary total energy with respect to the final binary total energy as a function of $b_{\rm max}$. The colours are the same as in Figs.~\ref{fig:cumulativ}, \ref{fig:bmax}, and~\ref{fig:Pmerg}.}
    \label{fig:dE}
\end{figure}

Finally, fig.~\ref{fig:dE} shows the relative total energy variation of the binaries, $dE/E=\left|E_{\rm fin}-E_{\rm in}\right|/\left|E_{\rm fin}\right|$, where $E_{\rm fin}$ and $E_{\rm in}$ are the total binary energy at the beginning and end of the simulation, respectively. The peak of the distribution shifts from larger to lower values of the relative total binary energy variation, when $k$ increases. For instance, the distributions for our fiducial model and for eq.~\ref{eq:otherb} with $k=1$ peak more than one order of magnitude above the distributions obtained for $k=4$ and $k=5$. In the former, only $2.6\%$ and $2.7\%$ of the binary systems yield $dE/E<10^{-2}$, while this number increases to $20\%$ and $31\%$ for the latter. 

\end{appendix}

\end{document}